\documentclass[pre,aps,reprint,superscriptaddress]{revtex4-1}
\usepackage[T1]{fontenc}
\usepackage[latin9]{inputenc}
\usepackage{color}
\usepackage{float}
\usepackage{amsmath}
\usepackage{amssymb}
\usepackage{graphicx}

\makeatletter

\@ifundefined{textcolor}{}
{%
  \definecolor{BLACK}{gray}{0}
  \definecolor{WHITE}{gray}{1}
  \definecolor{RED}{rgb}{1,0,0}
  \definecolor{GREEN}{rgb}{0,1,0}
  \definecolor{BLUE}{rgb}{0,0,1}
  \definecolor{CYAN}{cmyk}{1,0,0,0}
  \definecolor{MAGENTA}{cmyk}{0,1,0,0}
  \definecolor{YELLOW}{cmyk}{0,0,1,0}
}

\begin{document}

\title{Practical Guide to Quantum Phase Transitions in Quantum-Dot-Based Tunable Josephson Junctions}

\author{A. Kadlecov\'a}
\affiliation{Department of Condensed Matter Physics, Faculty of Mathematics and
Physics, Charles University in Prague, Ke Karlovu 5, CZ-121 16 Praha
2, Czech Republic}

\author{M. \v{Z}onda}
\affiliation{Institute of Physics, Albert Ludwig University of Freiburg, Hermann-Herder-Strasse
3, 791 04 Freiburg, Germany}
\affiliation{Department of Condensed Matter Physics, Faculty of Mathematics and
Physics, Charles University in Prague, Ke Karlovu 5, CZ-121 16 Praha
2, Czech Republic}

\author{V. Pokorn\'y}
\affiliation{Department of Condensed Matter Physics, Faculty of Mathematics and
Physics, Charles University in Prague, Ke Karlovu 5, CZ-121 16 Praha
2, Czech Republic}

\author{T. Novotn\'y}
\email{tno@karlov.mff.cuni.cz}
\affiliation{Department of Condensed Matter Physics, Faculty of Mathematics and
Physics, Charles University in Prague, Ke Karlovu 5, CZ-121 16 Praha
2, Czech Republic}

\date{\today}
\begin{abstract}
Quantum dots attached to BCS superconducting
leads exhibit a $0-\pi$ impurity quantum phase transition, which
can be experimentally controlled either by the gate voltage or by
the superconducting phase difference. For the pertinent superconducting 
single-impurity Anderson model, we newly present two simple
analytical formulae describing the position of the phase boundary
in parameter space for the weakly correlated and Kondo regime, respectively.
Furthermore, we show that the two-level approximation provides an
excellent description of the low temperature physics of superconducting
quantum dots near the phase transition. We discuss reliability and
mutual agreement of available finite temperature numerical methods
(Numerical Renormalization Group and Quantum Monte Carlo) and suggest
a novel approach for efficient determination of the quantum phase
boundary from measured finite temperature data. Our results enable fast and efficient, 
yet reliable characterization and design of such nanoscopic tunable Josephson junction devices.     
\end{abstract}
\maketitle

\section{Introduction}

Low temperature nanostructures involving quantum dots attached to
superconductors have been intensively studied in the past two decades
--- see Refs.~\cite{Rodero-2011} for theoretical and \cite{Wernsdorfer-2010}
for experimental overviews. A number of various setups involving several
superconducting and/or normal leads have been thus far realized using
a variety of systems (single molecules such as $C_{60}$, carbon nanotubes,
semiconducting InAs nanowires etc.) as the central functional element
(quantum dot) \cite{Morpurgo-1999,Kasumov-1999,Kasumov-2003,Jarillo-2006,vanDam-2006,Jorgensen-2006,Cleuziou-2006,Jorgensen-2007,Grove-2007,Pallecchi-2008,Zhang-2008,Jorgensen-2009,Liu-2009,Eichler-2009,Winkelmann-2009,Pillet-2010,Katsaros-2010,Maurand-2012,Lee-2012,Pillet-2013,Kumar-2014,Delagrange-2015,Delagrange-2016,Xu-2017,Delagrange-2018-PhysicaB, Farinacci-2018}.
Parameters of such systems are typically tunable by gate voltage,
which changes the single-particle energies on the dot, and in case
of SQUID setups by the magnetic flux through the loop tuning the phase
difference across these generalized Josephson junctions. Their envisioned applications 
range from various sensors and detectors (e.g., single-molecule SQUIDs \cite{Cleuziou-2006, Bouchiat-2009}) to building blocks of quantum information technologies \cite{Wernsdorfer-2010}.

One of the simplest setups involves a quantum dot attached to just
two superconducting leads whose relative superconducting phase difference
$\text{\ensuremath{\varphi}}$ can be tuned leading to the flow of
the Josephson supercurrent through the junction. Very often such a
system can be even quantitatively described by the single impurity
Anderson model (SIAM) coupled to BCS leads \cite{Luitz-2012}, which
exhibits an impurity quantum phase transition. This so called $0-\pi$
transition corresponds to the change of the system ground state from
a non-magnetic singlet to a spin-degenerate doublet and is accompanied
by the sign-change of the supercurrent (from positive in the $0$-phase
to negative in the $\pi$-phase) \cite{vanDam-2006,Cleuziou-2006,Jorgensen-2007,Eichler-2009,Maurand-2012,Delagrange-2015,Delagrange-2016,Xu-2017,Delagrange-2018-PhysicaB}
and crossing of the Andreev bound states (ABSs) at the Fermi energy
\cite{Pillet-2010,Pillet-2013,Chang-2013,Xu-2017}. Depending on the
relative strength of the on-dot Coulomb interaction the $0$-phase
ground state singlet can be predominantly BCS-like (for weak interaction)
or Kondo-like (strong correlations) with a broad crossover between
these two limiting cases. This physical picture has been firmly established
over the years by various analytic and numeric theoretical methods
\cite{Matsuura-1977,Glazman-1989,Rozhkov-1999,Yoshioka-2000,Siano-2004,Choi-2004,Sellier-2005,Novotny-2005,Karrasch-2008,Meng-2009,Luitz-2012,vonOppen-2017}
and fully qualitatively confirmed already by pilot experiments \cite{vanDam-2006,Cleuziou-2006,Jorgensen-2007}.

However, recent experiments using the SQUID setup allowing a high
level of tunability \cite{Delagrange-2015,Delagrange-2016,Xu-2017,Delagrange-2018-PhysicaB}
have revealed difficulties involved in making a quantitative comparison
with theory. Heavy numerical tools such as the Quantum Monte Carlo
(QMC) or Numerical Renormalization Group (NRG) turn out to be too
costly as for the computational resources to allow for broader scans
throughout the model parameter space, which are necessary for an efficient
and reliable identification of the experimental situation. They seem
to be quite inconvenient for the initial phase of the data analysis,
which should place the given experimental setup into the proper context
of rough parameter values, and for capturing the global trends induced
by coarse-grained parameter changes. 

This task rather calls for a simple, ideally analytical or very efficient
numerical technique which would parse the parameter space grossly.
As a next step more elaborate methods including QMC and/or NRG could
be used to fine-tune the parameters, yet taking into account the common
experimental accuracy of 10-20\%, quite often these precise methods
may not be required at all. Here, we offer two simple analytical formulae
for the position of the $0-\pi$ phase boundary in the complementary
weakly interacting and strongly correlated (Kondo) regimes, respectively.
They are based on the combination of analytical insights and NRG data
and with a reasonable precision cover a big part of the SIAM parameter
space. 

Another issue concerns finite temperatures: the phase boundary is
a ground-state, i.e. zero-temperature quantity but the experiments
are naturally performed at finite (even if ideally very small) temperatures.
The task of extrapolating to zero-temperature from finite temperature
experimental data is principally nontrivial and, as we will show,
it has not been so far addressed properly. We identify a very simple
and straightforward method how to extract zero-temperature quantities
directly from finite-temperature data without the need for any post-processing. 

\section{Model and notation }
\begin{subequations}
As explained above we consider the single-impurity Anderson model
of a quantum dot connected to two BCS superconducting leads. The full
Hamiltonian reads 
\begin{align}
\mathcal{H} & =\mathcal{H}_{{\rm dot}}+\sum_{\alpha}(\mathcal{H}_{{\rm lead}}^{\alpha}+\mathcal{H}_{T}^{\alpha}),\label{eq:Hamiltonian_1}
\end{align}
where $\alpha=L,R$ denotes the left and right superconducting leads.
The dot Hamiltonian 
\begin{equation}
\mathcal{H}_{{\rm dot}}=\varepsilon\sum_{\sigma=\uparrow,\downarrow}d_{\sigma}^{\dagger}d_{\sigma}+Ud_{\uparrow}^{\dagger}d_{\uparrow}d_{\downarrow}^{\dagger}d_{\downarrow}
\end{equation}
describes an impurity with the spin-degenerate single-particle level
$\varepsilon$ and the local Coulomb interaction $U$ in case of the
doubly occupied dot. Operators $d_{\sigma}^{\dagger}$ ($d_{\sigma}$)
create (annihilate) on-dot electrons with spin $\sigma$. The BCS
Hamiltonian of the superconducting leads is 
\begin{equation}
\mathcal{H}_{{\rm lead}}^{\alpha}=\sum_{\mathbf{k}\sigma}\varepsilon_{\alpha}(\mathbf{k})\,c_{\alpha\mathbf{k}\sigma}^{\dagger}c_{\alpha\mathbf{k}\sigma}-\Delta_{\alpha}\sum_{\mathbf{k}}(e^{i\varphi_{\alpha}}c_{\alpha\mathbf{k}\uparrow}^{\dagger}c_{\alpha\mathbf{\,-k}\downarrow}^{\dagger}+\textrm{H.c.}),
\end{equation}
where $c_{\alpha\mathbf{k}\sigma}^{\dagger},\,c_{\alpha\mathbf{k}\sigma}$
are the creation and annihilation operators of electrons with momentum
$\mathbf{k}$ and spin $\sigma$, $\Delta_{\alpha}$ is the amplitude
of the superconducting gap in the lead $\alpha$, and $\varphi_{\alpha}$
is its superconducting phase. We denote by $\varphi\equiv\varphi_{L}-\varphi_{R}$
the phase difference between the two superconducting leads. The last
term in Eq.~\eqref{eq:Hamiltonian_1} is the tunnel coupling Hamiltonian
\begin{equation}
\mathcal{H}_{T}^{\alpha}=\sum_{\mathbf{k}\sigma}(t_{\alpha\mathbf{k}}c_{\alpha\mathbf{k}\sigma}^{\dagger}d_{\sigma}+\textrm{H.c.}),
\end{equation}
with $t_{\alpha\mathbf{k}}$ denoting the tunneling matrix elements.
We assume the tunnel-coupling magnitudes $\Gamma_{\alpha}(\varepsilon)\equiv\pi\sum_{\mathbf{k}}|t_{\alpha\mathbf{k}}|^{2}\delta(\varepsilon-\varepsilon_{\alpha}(\mathbf{k}))$
to be constant in the energy range of interest $\Gamma_{\alpha}(\varepsilon)\simeq\Gamma_{a}$.
\end{subequations}

The model is described by just a few parameters: the dot level energy
$\text{\ensuremath{\varepsilon}}$ (which can be experimentally tuned
by the gate voltage), the local Coulomb interaction between dot electrons
$U$, the total coupling strength $\Gamma\equiv\Gamma_{L}+\Gamma_{R}$
and the tunnel asymmetry of the setup $a=\Gamma_{L}/\Gamma_{R}$,
the phase difference $\varphi$ (which, if the junction is a part
of a SQUID, can be controlled by an applied magnetic field \cite{vanDam-2006,Cleuziou-2006,Maurand-2012,Delagrange-2015,Delagrange-2016,Xu-2017,Delagrange-2018-PhysicaB}),
and the superconducting gaps $\Delta_{\alpha}$. Throughout this whole
article we will assume the generic experimental situation of equal
gaps $\Delta_{L}=\Delta_{R}\equiv\text{\ensuremath{\Delta}}$, which
implies that we can use the symmetry-asymmetry relation discovered
in Ref.~\cite{Kadlecova-2017} to simplify the model by introduction
of the compact quantity 
\begin{equation}
\chi=\chi(\varphi,\,a)\equiv1-\frac{4a}{(a+1)^{2}}\sin^{2}\frac{\varphi}{2},\label{eq:defchi}
\end{equation}
on which the on-dot quantities (including especially the phase boundary)
exclusively depend, i.e.~the two parameters $a$ and $\varphi$ are
reduced to a single one $\chi$.

For the normal-state Kondo temperature we use the expression based
on Wilson's definition via magnetic susceptibility \cite{Wilson-1975,Haldane-78b}
\begin{equation}
k_{B}T_{K}\equiv0.29\sqrt{\Gamma U}\exp\left(-\frac{\pi|\varepsilon|(\varepsilon+U)}{2\Gamma U}\right).\label{eq:defTk}
\end{equation}
Eventually, we will be further using (when convenient) the shifted
and normalized level energy 
\begin{equation}
\tilde{\varepsilon}\equiv\frac{\varepsilon+U/2}{U/2}=1+\frac{2\varepsilon}{U},\label{eq:epsilontilda}
\end{equation}
a dimensionless number which is zero at half-filling ($\ensuremath{\varepsilon}=-U/2$). 

\section{Zero-temperature phase boundaries}

\begin{figure}
\includegraphics[width=1\columnwidth]{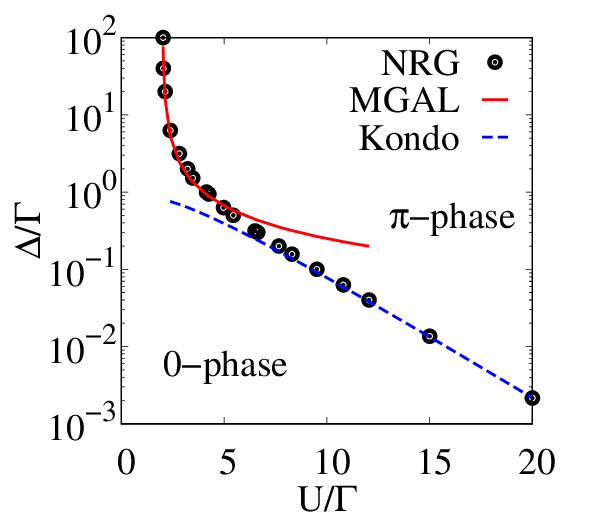}
\caption{Phase diagram for the superconducting SIAM at half-filling ($\varepsilon=-U/2$)
and $\chi=1\,(\varphi=0)$. We illustrate the ranges of validity of
the formulae given in Secs.~\ref{subsec:The-weakly-correlated-regime}
and \ref{subsec:Kondo-regime}. Black bullets represent Numerical
Renormalization Group data, the red line is the MGAL prediction \eqref{eq:mGAL},
and the blue dashed line corresponds to $\Delta_{C}\approx4.29T_{K}$
given by Eq.~\eqref{eq:Kondo} for $\chi=1$. 
\label{fig:Validity_Range}}
\end{figure}

For ground states, features of the system are known to be well captured
by the NRG. However, these computations can be time-consuming and
it is therefore advantageous to have other, possibly less precise
but significantly easier tools at hand. In two complementary limits
we have found simple analytical formulae which capture the position
of the $\text{0 -- \ensuremath{\pi}}$ phase boundary in the parameter
space. The ``MGAL'' approximation presented in Sec.~\ref{subsec:The-weakly-correlated-regime}
deals with the weakly correlated regime characterized by moderate
$U/\Gamma$ ratios. On the other hand, in Sec.~\ref{subsec:Kondo-regime}
we comment on the strongly-correlated Kondo regime of the quantum
dot, taking into account the $\text{\ensuremath{\chi\,(\varphi)}}$
dependence. Fig.~\ref{fig:Validity_Range} illustrates the ranges
of validity of our predictions. At half-filling ($\tilde{\varepsilon}=0$)
and for $\chi=1\,(\varphi=0)$ the MGAL approximation is valid up
to $U/\Gamma\lesssim5$. On the other hand, Kondo physics prevails
for $U/\Gamma\gtrsim7$. The intermediate range can be well and very
fast captured by the numerical solution of the second-order perturbation
theory of Refs.~\cite{Zonda-2015,Zonda-2016} (in particular, see Fig.~6
in Ref.~\cite{Zonda-2016}) for which we provide publicly accessible
code \cite{SQUAD-code}.

\subsection{Weakly-correlated regime\label{subsec:The-weakly-correlated-regime}}

\begin{figure}
\includegraphics[width=1\columnwidth]{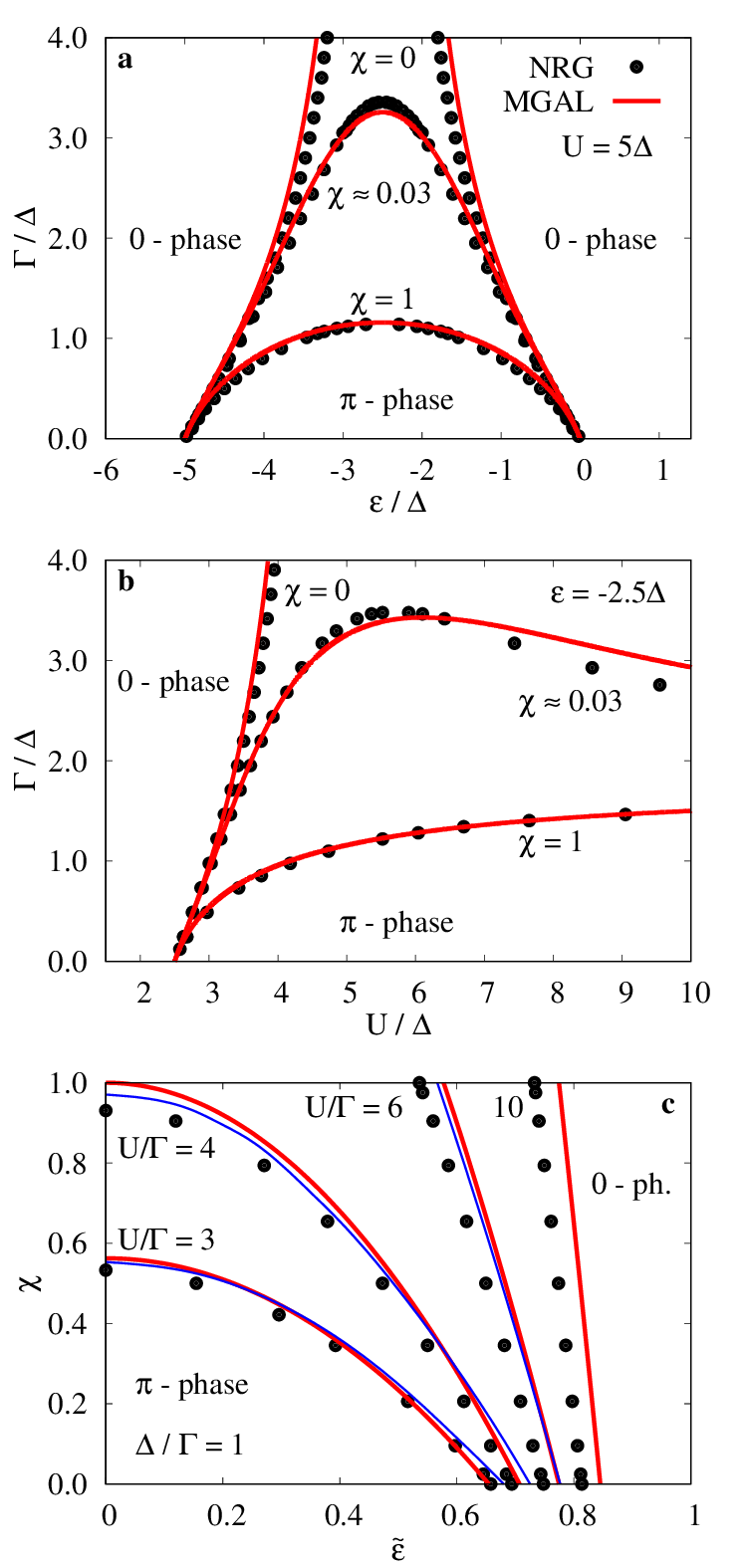}\caption{Zero-temperature phase diagrams of the superconducting quantum dot
in the weakly-correlated regime. Black bullets represent NRG data,
while the red line is the prediction by the MGAL formula \eqref{eq:mGAL}
and blue lines in panel c are the SOPT solutions. (a) $\text{\ensuremath{\Gamma} -- \ensuremath{\varepsilon}}$
dependence for a junction with $U=5\Delta$. The phase boundary is
shown for $\chi=1$, $\chi=0$ and a small but nonzero $\chi=0.0325$.
(b) $\Gamma-U$ dependence away from half filling ($\varepsilon=-2.5\Delta$).
(c) $\chi-\tilde{\varepsilon}$ dependence for different $U/\Gamma$
ratios ($U/\Gamma=$3, 4, 6 and 10) at $\Gamma=\text{\ensuremath{\Delta}}$.
\label{fig:mGAL_tests}}
\end{figure}

By analyzing NRG data obtained by the ``NRG Ljubljana'' code \cite{Ljubljana-code}
we have found (for more details see the Appendix) that for the weakly
correlated quantum dot regime the phase boundary can be approximated
with the equation
\begin{align}
\chi&=\mathcal{U}^{2}-\mathcal{U}(\mathcal{U}+1)\tilde{\varepsilon}^{2},\label{eq:mGAL}
\intertext{where} 
\mathcal{U}&\equiv\frac{U}{2\Gamma}\frac{\Delta}{\Gamma+\Delta}
\end{align}
and $\chi=\chi(\varphi,\,a),\,\tilde{\varepsilon}$ are given by Eqs.~\eqref{eq:defchi}
and \eqref{eq:epsilontilda}, respectively. For $\chi=1\,(\varphi=0)$
the relation \eqref{eq:mGAL} reduces to $1-\tilde{\varepsilon}^{2}=\tfrac{1}{\mathcal{U}}$.
We call Eq.~\eqref{eq:mGAL} the \emph{Modified Generalized Atomic
Limit} (MGAL), referring to the previously derived Generalized Atomic
Limit (GAL) approximation \cite{Zonda-2015,Zonda-2016} which is identical
to MGAL at the half-filling $\tilde{\varepsilon}=0$.

To illustrate the agreement of Eq.~\eqref{eq:mGAL} with the NRG
data, we present zero-temperature phase diagrams for different parameter
sub-spaces in Fig.~\ref{fig:mGAL_tests}, namely the $\Gamma-\varepsilon$
phase diagram in \ref{fig:mGAL_tests}(a), the $\Gamma-U$ diagram
away from half-filling in \ref{fig:mGAL_tests}(b) and, finally, several
phase-transition boundaries in the $\chi-\tilde{\varepsilon}$ ($\varphi-\varepsilon$)
plane in Fig.~\ref{fig:mGAL_tests}(c). Eq.~\eqref{eq:mGAL} is
mostly in a pretty good agreement with the NRG and significantly outperforms
previously-known analytic formulas including the atomic limit \cite{Karrasch-2008,Bauer-2007},
Hartree-Fock prediction, and the GAL away from half-filling \cite{Zonda-2015,Zonda-2016}.
We therefore suggest it as a simple first estimate of the position
of the phase boundary in the weakly-correlated regime (cf.~Fig.~\ref{fig:Validity_Range}).

A more elaborate method of determining the phase boundary in the weakly-correlated
regime is the second-order perturbation theory (SOPT) \cite{Zonda-2015,Zonda-2016}.
This method is based on the perturbation expansion technique in the
Coulomb interaction $U$. Although this method is unable to describe
the $\pi$-phase due to its double-degenerate ground state, it provides
reliable description of the $0$-phase including its phase boundary
up to $U/\Gamma\approx10$ (not too far from half-filling), see Fig.~\ref{fig:mGAL_tests}c.
This method is numerical and, consequently, it is much harder to implement
than analytical MGAL, nevertheless an efficient, free, and easy-to-use
Python code called SQUAD is available~\cite{SQUAD-code}. Numerical
perturbative calculations are orders-of-magnitude faster than the
NRG code and constitute a very reasonable compromise between the accuracy
and numerical costs in the weak-to-moderately correlated regime. For
all possible aspects and details of the perturbation theory we refer
the reader to Ref.~\cite{Zonda-2016}.

\subsection{Kondo regime \label{subsec:Kondo-regime}}

In the Kondo regime the phase boundary is widely believed to be a
universal function of $T_{K}/\Delta$ and specifically to occur at
$T_{K}\approx\Delta$. In Ref.~\cite{Kadlecova-2017} we argued that
the coupling asymmetry $a$ must play some role, however, we left
the question of universality open. This section establishes that the
phase boundary can indeed be described by a universal function of
$T_{K}/\Delta$ if $\chi=\chi(\varphi,\,a)$ given by Eq.~\eqref{eq:defchi}
is used as a variable.

\begin{figure*}[!htbp]
\includegraphics[width=1\textwidth]{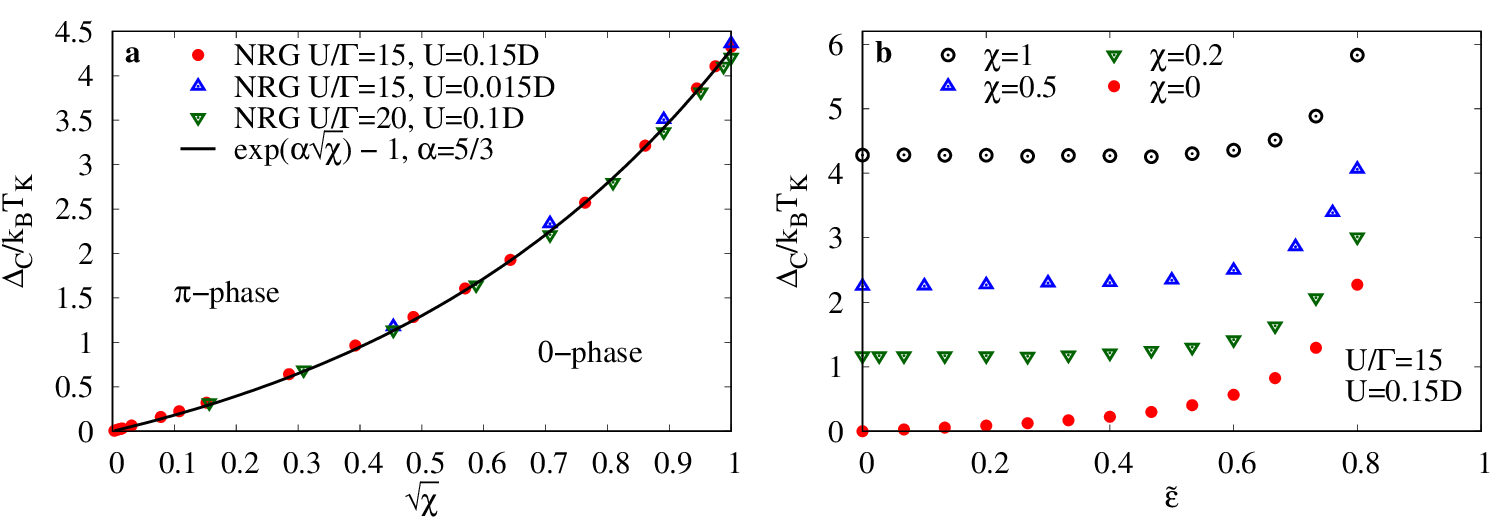}

\caption{(a) Universal shape in the Kondo regime of the ratio of the critical
value of the gap $\Delta_{C}$ over the Kondo temperature $T_{K}$
\eqref{eq:defTk} as a function of variable $\chi$ \eqref{eq:defchi}.
Points represent NRG data with $U/\Gamma=15,\:20$ and different
values of the bandwidth $D$. The solid line corresponds to $\Delta_{C}/k_{B}T_{K}=\exp(\alpha\sqrt{\chi})-1$
with $\alpha=5/3$. (b) The $\text{\ensuremath{\tilde{\varepsilon}}}$
dependence of $\Delta_{C}/T_{K}$ for $\text{\ensuremath{\chi}=0,\,0.2,\,0.5, and 1}$.
\label{fig:Delta/Tk} }
\end{figure*}

The formula for the critical value of the gap $\Delta_{C}$ determined from the NRG data {[}Fig. \ref{fig:Delta/Tk}(a){]}
and valid for our definition of $T_{K}$ \eqref{eq:defTk} (if another convention is
used, the formula should be properly rescaled) reads

\begin{equation}
\frac{\Delta_{C}}{k_{B}T_{K}}=\exp\left(\alpha\sqrt{\chi}\right)-1,\label{eq:Kondo}
\end{equation}
where $\alpha\approx5/3$. More accurately, we have fitted three different
sets of numerical data {[}shown in Fig.~\ref{fig:Delta/Tk}(a){]}
and we have found that $\alpha_{1}=1.65\pm0.02$ for $U/\Gamma=20$,
$U=0.1D$, $\alpha_{2}=1.67\pm0.02$ for $U/\Gamma=15$, $U=0.15D$,
and $\alpha_{3}=1.69\pm0.03$ for $U/\Gamma=15$, $U=0.015D$, where
$D$ is the bandwidth used in the NRG calculations. Ideally the calculation
should be performed in the limit of an infinite band, hence the (necessary)
choice of a finite $D$ influences the numerical results slightly.

The dependence in Fig.~\ref{fig:Delta/Tk}(a) was calculated at half-filling,
$\tilde{\varepsilon}=0$. Fig.~\ref{fig:Delta/Tk}(b) reveals that
$\tilde{\varepsilon}$ dependence is very weak up to $\tilde{\varepsilon}\approx0.4$,
significantly departing from the value predicted by \eqref{eq:Kondo}
for $\tilde{\varepsilon}\approx0.6$. For $\chi=0$, which can only
be achieved for $\varphi=\pi$ in a perfectly symmetric junction with
$a=1$ \cite{Kadlecova-2017}, and exactly at half-filling, there
is no phase transition, but a small critical gap is found with any
departure from half-filling \footnote{Note that the Kondo temperature \eqref{eq:defTk} is also an approximation
valid around $\tilde{\varepsilon}\approx0$.}. Results in Fig.~\ref{fig:Delta/Tk}(b) are in agreement with Ref.~\cite[Fig. 9a]{Yoshioka-2000}.
Authors of this previous study have tested the $\varepsilon$-independence
for two different values of $U/\text{\ensuremath{\Gamma}}$ and concluded
that the universality breaks down in the valence fluctuation regime
$|\varepsilon|\lesssim\pi\Gamma$ ($|\tilde{\varepsilon}|\gtrsim1-2\pi\Gamma/U\approx0.58$).

As given by Eq.~\eqref{eq:Kondo}, for $\chi=1$ (corresponding to
$\varphi=0$) the phase transition appears (for our definition of
$T_{K}$ and $\alpha=5/3$) at $\Delta_{C}/T_{K}\approx4.29$. For
any nonzero $\varphi$ the critical gap will be smaller. 

\section{Finite temperatures }

In superconducting quantum dot devices the $0-\pi$ transition reflects
an underlying impurity quantum phase transition between the singlet
and doublet ground states, a crossing of the two lowest-energy many-body
levels. At zero temperature, the quantum critical point (QCP) is clearly
signaled by a jump in the supercurrent and the change of its sign,
however with increasing finite temperature the current-phase relation (CPR) becomes continuous and the point where the supercurrent
changes sign shifts away from the QCP. This complicates the determination
of the position of the QCP from real experimental data, as well as
from the results of strictly finite-temperature numerical methods
such as QMC. In Sec.~\ref{subsec:Low-temperature-physics}, we present
a simple physical argument that the crossing-point of the finite temperature
current-phase relations coincides with the QCP at low enough temperatures.
Moreover, the crossing can be observed not only for the current as
a function of the phase difference but basically for any physical
quantity as a function of any parameter that induces the singlet-doublet
phase transition. We further discuss the temperature range of applicability
of the underlying two-level approximation and why previously used
methods of estimating the QCP from the zero-crossing of the Josephson
current lead to inaccurate results (Sec.~\ref{subsec:Determining-the-QCP}). 
The two-level approximation expressed in Eq.~\eqref{eq:mix} does not only hold for our system but is universally applicable to impurity quantum phase transitions of the first order regardless of their physical realization and microscopic origin.    

Finite-temperature results can be obtained by two complementary numerically
exact methods, namely the NRG and QMC. NRG is a reliable method for
the ground state properties. It can also provide trustworthy results
for low enough temperatures but the high ones are usually beyond its
scope. On the other hand, the QMC is ideal for high temperatures but
its computational demands rapidly increase with decreasing temperature.
For quantum dots, there is a temperature range where both NRG and
QMC are commonly used, but accuracy of both finite temperature NRG
and low-temperature QMC is sometimes subject to questions. Also, while
for single quantum dots such as our system NRG is generally less computationally
demanding than QMC, for more complicated setups such as multiple quantum
dots or dots connected to multiple terminals QMC quickly becomes the
method of choice. It is therefore highly desirable to establish whether
these two methods are in agreement for systems where their ranges
of applicability overlap. Therefore, we have tested compatibility
of both methods for our finite-temperature data.

\begin{figure*}
\includegraphics[width=0.75\textwidth]{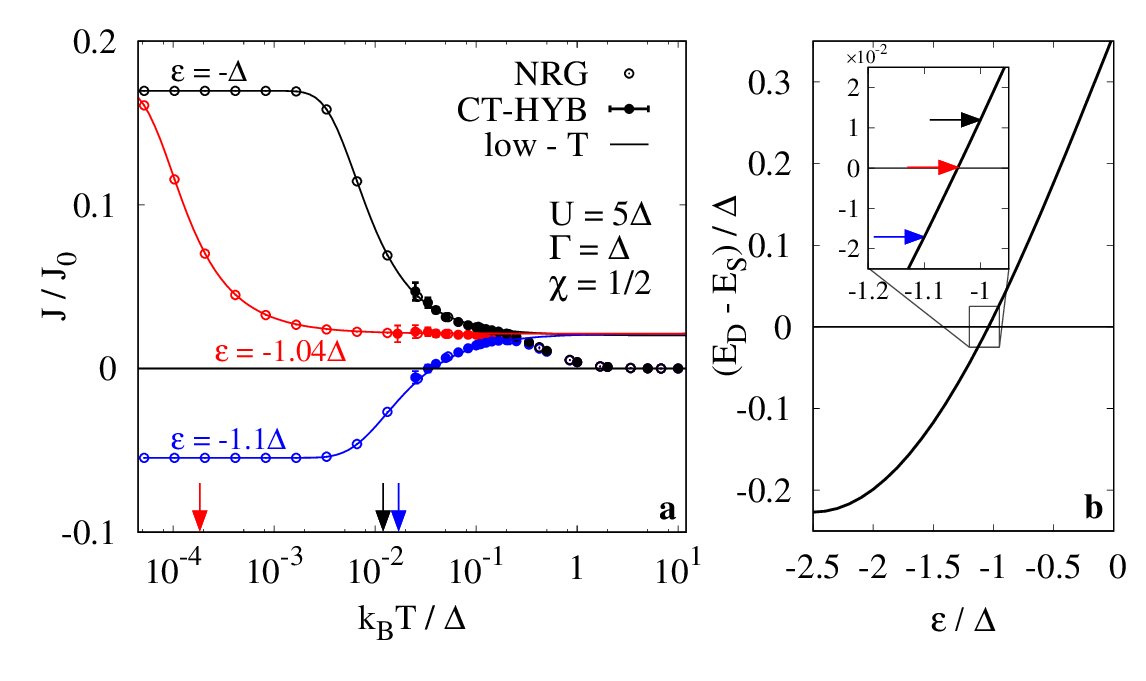}\caption{(\textbf{a}) Low-temperature behavior of the normalized Josephson
current ($J_{0}\equiv2e\Delta/\hbar$) for $U=5\Delta$, $\Gamma=\Delta$
and $\chi=1/2$. Empty circles with points represent NRG data,
full circles with error bars are CT-HYB results. The two methods agree
within the QMC error bars. Solid lines correspond to the low-temperature
prediction of Eq. \eqref{eq:mix} in the main text. The three lines
correspond to different values of the energy level $\varepsilon$
close to the phase boundary, so that the black line ($\varepsilon=-\Delta$)
shows low temperature behavior of the system in the $0$-phase, the
blue line ($\varepsilon=-1.1\Delta$) in the $\pi$-phase, and the
red one ($\varepsilon=-1.04\Delta$) is just slightly above the critical
value. Arrows mark corresponding absolute values of zero-temperature
Andreev bound states energies. (\textbf{b}) Difference between the
singlet ($E_{S}$) and doublet ($E_{D}$) ground state energy, which
corresponds to the energy of the Andreev bound state. The quantum
phase transition takes place at $E_{D}=E_{S}$. Inset: detail around
$E_{D}=E_{S}$. Arrows mark energies corresponding to those in
panel (\textbf{a}).\label{fig:plat}}
\end{figure*}

In our calculations we have used finite-temperature NRG from the ``NRG
Ljubljana'' code \cite{Ljubljana-code}, while QMC has been done
using the TRIQS/CTHYB continuous-time hybridization-expansion solver
\cite{cthyb2016}. The superconducting pairing is introduced to the
QMC method using a canonical particle-hole transformation in the spin-down
sector, mapping the system to an impurity Anderson model with attractive
interaction \cite{Luitz-2010,Pokorny-2018}. The comparisons of the
two methods are shown in Figs.~\ref{fig:plat}(a) and \ref{fig:Luitz},
where in the overlapping temperature range the NRG and QMC data coincide
within the QMC error bars. The agreement implies that both methods
are reliable for the experimentally-relevant range of temperatures.

\subsection{Low-temperature physics: two-level approximation \label{subsec:Low-temperature-physics}}

For low temperatures, the lowest (many-body) energy levels of a system
become most significant. Due to the superconducting gap of single-particle
excitations in our system, the lowest-lying states are discrete. In
the spin-degenerate case (without external magnetic field) considered
here there may be one or two discrete excited states below the single-particle
continuum starting at the gap. We are mainly interested in the vicinity
of the QCP where just one of these discrete excited states exchanges
its role with the ground state (one of these two is a singlet and
the other doublet). The other excited state, if it exists as a discrete
state, is much higher in energy and can be neglected together with
the continuum. We will now formalize and show some consequences of
this idea.

Starting with the canonical average $\bar{X}\equiv\frac{1}{Z}\sum_{i}X_{i}\exp(-\beta E_{i})$
of an observable $X$, we explore the low temperature regime $k_{B}T\ll\Delta$.
As discussed above we can approximate the sum by taking the two lowest-energy
states only. We obtain

\begin{equation}
X(y,\,T)\simeq\frac{X_{S}(y)e^{-\beta E_{S}(y)}+2X_{D}(y)e^{-\beta E_{D}(y)}}{e^{-\beta E_{S}(y)}+2e^{-\beta E_{D}(y)}},\label{eq:mix}
\end{equation}
where $X_{S(D)}$ is the zero-temperature value of the observable
in the singlet (doublet) state, $E_{S(D)}$ is the associated energy
of the singlet (doublet; factor $2$ reflects its twofold degeneracy)
state, $y$ is any model parameter (e.g., the phase difference $\varphi$)
and $\beta\equiv1/k_{B}T$. Note that the fraction can be reduced
by $e^{-\beta E_{S}(y)}$ to let it depend only on the energy difference
corresponding to the energy of the Andreev bound states (ABS), $E_{\mathrm{ABS}}(y)\equiv E_{D}(y)-E_{S}(y)$.

To illustrate the physics of Eq.~\eqref{eq:mix}, in Fig.~\textcolor{black}{\ref{fig:plat}(a)}
we present the dependence of the supercurrent on temperature for three
chosen values of $\varepsilon$ from the vicinity of the phase transition.
The empty circles with points have been calculated with the NRG, while
full circles with error bars represent the QMC results (for more specification
and comparison of the methods see the discussion just above this subsection).
The solid lines show the prediction of Eq.~\eqref{eq:mix} with zero-temperature
values of $J_{S(D)}(\varepsilon)$ and $E_{\mathrm{ABS}}(\varepsilon)$
obtained by the NRG. They belong to $\text{\ensuremath{\varepsilon}}$
above, bellow, and very close to the critical value as shown in the
inset in panel (b), where the zero-temperature normalized energies
of the Andreev bound states $E_{\mathrm{ABS}}(\varepsilon)/\Delta$
are marked by arrows of the corresponding color. We see that the lines
start as near-constants in temperature at the value $J_{S(D)}(\varepsilon)$
for $k_{B}T\lesssim E_{\mathrm{ABS}}(\varepsilon)$ and approach $\left(J_{S}(\varepsilon)+2J_{D}(\varepsilon)\right)/3$
for $k_{B}T\gtrsim E_{\mathrm{ABS}}(\varepsilon)$ with the crossover
happening at $k_{B}T\approx E_{\mathrm{ABS}}(\varepsilon)$ (arrows
on the horizontal axis). In all cases, Eq.~\eqref{eq:mix} captures
perfectly the low-temperature behavior up to $k_{B}T\approx0.2\Delta$.
For even higher temperatures, the continuum of excitations above the
gap $\Delta$ comes into play and the two-level approximation \eqref{eq:mix}
necessarily breaks down.

\begin{figure*}
\includegraphics[width=1\textwidth]{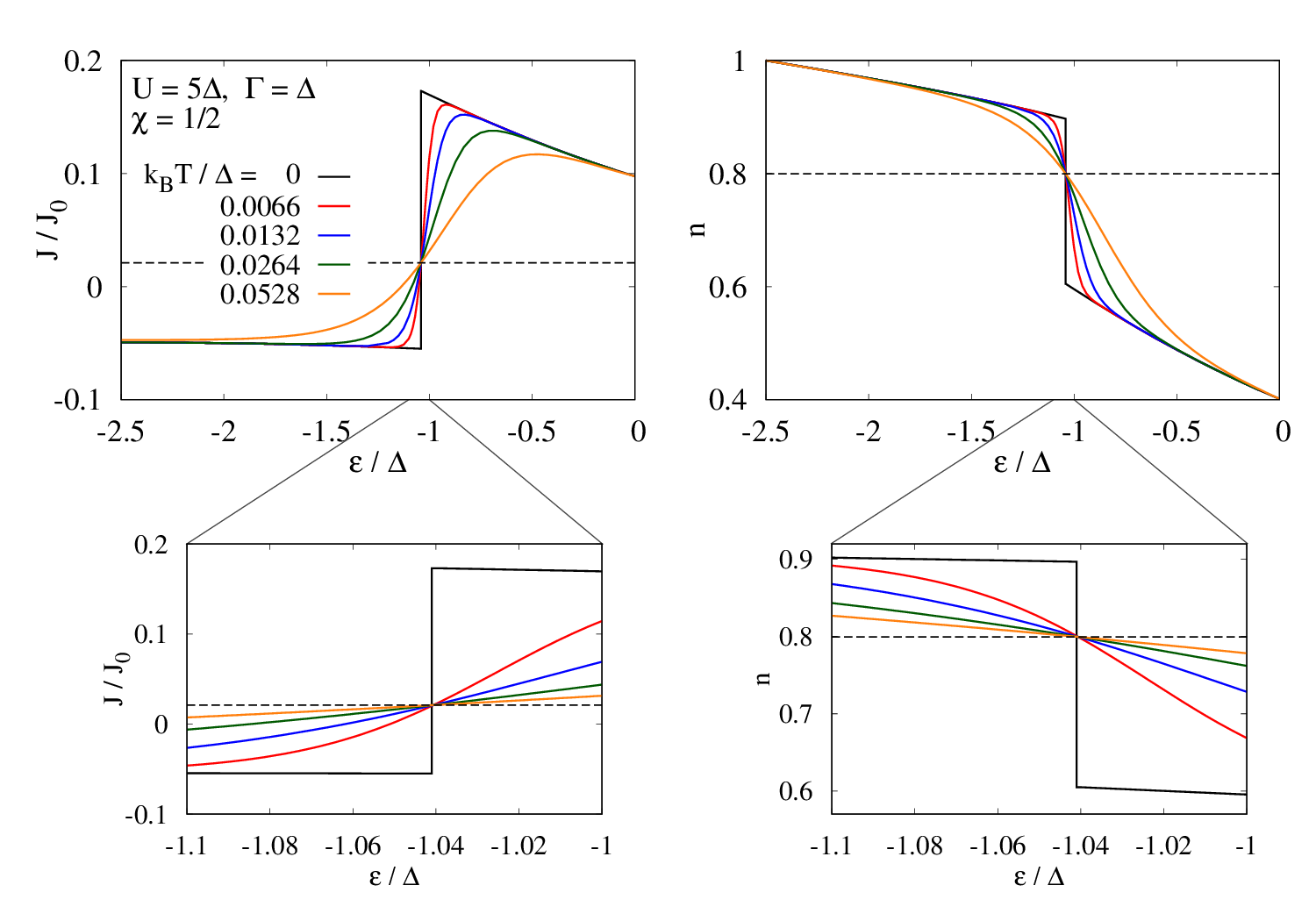}\caption{The normalized supercurrent (left panel; $J_{0}\equiv2e\Delta/\hbar$)
and the average dot occupation (right panel) as functions of the energy
level $\varepsilon$ for $U=5\Delta$, $\Gamma=\Delta$, $\chi=1/2$
and five values of temperature (in units of $\Delta$). Horizontal dashed
lines mark the values at the critical point predicted by Eq.~\eqref{eq:aveg},
$J=[J_{S}+2J_{D}]/3$ and $n=[n_{S}+2n_{D}]/3$, respectively. Bottom
panels are enlargements in the vicinity of the crossing points. The curves
are splines of the NRG data.\label{fig:FiniteT_test1}}
\end{figure*}

Exactly at the QCP the singlet and doublet many-body states cross,
meaning $E_{S}(y_{C})=E_{D}(y_{C})$. Consequently, from Eq.~\eqref{eq:mix}
we get the simple relation 
\begin{equation}
X(y_{C},\,T)=\frac{X_{S}(y_{C})+2X_{D}(y_{C})}{3},\label{eq:aveg}
\end{equation}
which does not depend on temperature (within the low-temperature regime
$k_{B}T\lesssim0.2\Delta$ justifying the two-level approximation).
We show a precise test of formula \eqref{eq:aveg} with data obtained
by the finite-temperature NRG in Fig.~\ref{fig:FiniteT_test1}. The
supercurrent (left panel) and average dot occupation (right panel)
are plotted as functions of $\varepsilon$ for five values of temperature.
The enlargements in the vicinity of the phase transition point prove that,
indeed, at this point all lines cross and have the value determined
by Eq.~\eqref{eq:aveg} (denoted by the horizontal dashed line). Although numerical
evidence that the crossing of finite-temperature current-phase relations
coincides with the QCP has been presented before (cf.~Refs.~\cite[Fig. 10]{Karrasch-2008}
and \cite[Fig. 1]{Choi-2005comm}), as far as we are aware the relevant
underlying physical mechanism expressed by Eq.~\eqref{eq:aveg} hasn't
been explicitly discussed yet.

\subsection{Determining the QCP from finite-temperature data \label{subsec:Determining-the-QCP}}

As Sec.~\ref{subsec:Low-temperature-physics} shows (Eq.~\eqref{eq:aveg}
and Fig.~\ref{fig:FiniteT_test1}), the crossing of different temperature
current phase relations may be a convenient way to straightforwardly
determine the position of the QCP from finite-temperature data. However,
the assumption $k_{B}T\ll\Delta$ used in our derivation may seem
limiting and, therefore, we have tested this method for parameters
that reflect a real experimental setup from Ref.~\cite{Delagrange-2015}.
Namely, in Fig.~\ref{fig:Luitz} we have recalculated the example
presented in the supplemental material of Ref.~\cite{Delagrange-2015}
with parameters reading $\Delta=0.17$ meV, $U=19\Delta$, $\Gamma_{L}+\Gamma_{R}=2.6\Delta$,
$a=\Gamma_{L}/\Gamma_{R}=4$, $\varepsilon=-4.8\Delta$, and the temperature
of the experiment $T_{\mathrm{exp}}=0.076\Delta/k_{B}$ ($150$~mK).
The upper panel of Fig.~\ref{fig:Luitz} reveals that the crossing works
up to at least $T=0.21\Delta/k_{B}$ ($420$~mK) analogously to the
findings of the previous subsection. This should leave enough room
for measuring a second dataset at a sufficiently higher temperature
to yield another well-distinguished CPR curve, so that the position
of the QCP could be read off directly from the intersection of the
experimental data without any need for post-processing. 
\begin{figure}
\includegraphics[width=1\columnwidth]{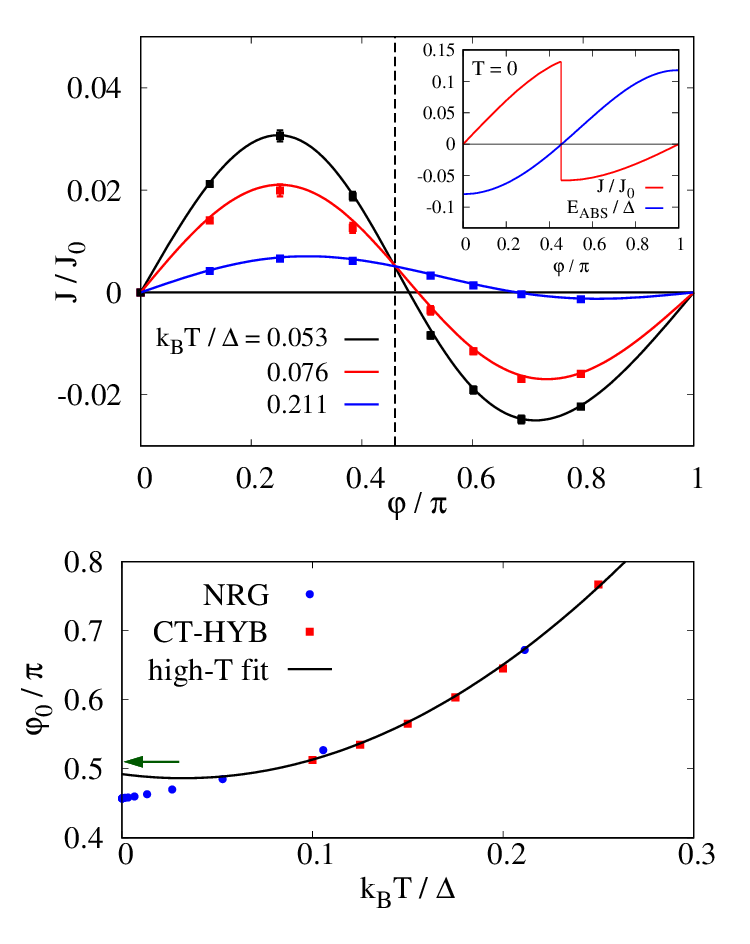}
\caption{Top: current-phase relations ($J_{0}\equiv2e\Delta/\hbar$)
for experimental parameters taken from the supplemental material of
Ref.~\cite{Delagrange-2015}. The points with error bars were obtained
using CT-HYB, solid lines represent the NRG results. The dashed line
marks the jump in the zero-temperature NRG result for the supercurrent
(see inset). The CPRs for different temperatures intersect at one
point. Inset: zero-temperature supercurrent and energy of the Andreev
bound state. Bottom: temperature dependence of the
zero of the current-phase relation, $\varphi_{0}(T)$, determined
by the condition $J(\varphi_{0}(T),T)=0$. Blue circles represent
NRG results, red squares CT-HYB data. The black solid line is a quadratic
fit of the high-temperature CT-HYB data in analogy with the method
used in Ref.~\cite{Delagrange-2015}. The green arrow denotes the
result of Ref.~\cite{Delagrange-2015}. The discrepancy between the
two calculations is caused by slightly different extrapolation procedures
as explained in the main text. \label{fig:Luitz}}
\end{figure}

Apart from being an unnecessary computational burden, the post-processing
itself might introduce an extra error into the interpretation of the
experimental data as we will now demonstrate on the method used in
Ref.~\cite{Delagrange-2015} to determine the QCP. In the supplemental
material the authors describe the procedure used for extracting the
critical phase difference $\varphi_{C}$ from the finite-temperature
QMC data. Their numerical calculations were performed using the continuous-time,
interaction-expansion (CT-INT) algorithm~\cite{Luitz-2010}. Few
data points for each CPR $J(\varphi,T)$ for various temperatures
between $145$ to $580$~mK were calculated and approximated
by a three-term Fourier series $I(\varphi)=a_{1}\sin(\varphi)+a_{2}\sin(2\varphi)+a_{3}\sin(3\varphi)$.
The critical phase difference $\varphi_{C}$ was then extrapolated
from the zeroes $\varphi_{0}(T)$ of these Fourier fits for various
finite temperatures using quadratic extrapolation (i.e., parabolic
fit $\varphi_{C}-\varphi_{0}(T)\propto T^{2}$) down to $T=0$. As
the result lies very close to the zero of the measured CPR for the
lowest experimental temperature $150$~mK, this value was taken as
the correct zero-temperature limit and thus the true critical phase.

However, our findings contradict such a conclusion. We have recalculated
the CPRs using CT-HYB algorithm with more attention given to the vicinity
of the zero-crossing points $J(\varphi,T)=0$ and performed the same
quadratic extrapolation, obtaining very similar results \footnote{The reason why our zero-temperature extrapolation does not coincide
precisely with the previous calculation (marked by the green arrow)
is most probably the Fourier fitting which we could avoid.} shown by the black line in the bottom panel of Fig.~\ref{fig:Luitz}.
Although this procedure seems perfectly plausible, we see that $\varphi_{C}$
obtained this way disagrees with the zero-temperature NRG result,
which nevertheless coincides with the aforementioned crossing of the
current-phase relations as it should.

To understand why the extrapolation method described in supplemental
material of Ref.~\cite{Delagrange-2015} failed to predict the correct
position of the QCP, we perform the low-temperature expansion of the supercurrent
using the two-level approximation \eqref{eq:mix}. Using the condition
$J(\varphi_{0}(T),\,T)=0$, from Eq.~\eqref{eq:mix} for
the intersection point we get $\varphi_{0}$: $J_{S}(\varphi_{0})+2J_{D}(\varphi_{0})e^{-\beta[E_{D}(\varphi_{0})-E_{S}(\varphi_{0})]}=0$.
We assume that the temperature is low enough so that $\varphi_{0}(T)$
is in the close vicinity of $\varphi_{C}$ and can be replaced by
it in the supercurrents $J_{S(D)}(\varphi_{0})\simeq J_{S(D)}(\varphi_{C})$.
Moreover, we perform a well-justified linear expansion of the ABS
energy in the exponent (see the blue curve in the inset of Fig.~\ref{fig:Luitz})
remembering that $E_{S}(\varphi_{C})=E_{D}(\varphi_{C})$ at the critical
point and $J(\varphi)\equiv\frac{2e}{\hbar}\frac{dE(\varphi)}{d\varphi}$
arriving at $E_{D}(\varphi_{0})-E_{S}(\varphi_{0})\simeq\frac{\hbar}{2e}[J_{D}(\varphi_{C})-J_{S}(\varphi_{C})]\left(\varphi_{0}-\varphi_{C}\right)$,
With these approximations we get a condition for the $\varphi_{0}(T)$
which reads

\begin{equation}
\varphi_{0}(T)\simeq\varphi_{C}+k_{B}T\,\frac{2e}{\hbar}\frac{\ln\left(2|J_{D}(\varphi_{C})|\right)-\ln J_{S}(\varphi_{C})}{J_{D}(\varphi_{C})-J_{S}(\varphi_{C})}\label{eq:linear_in_T}
\end{equation}
and, most importantly, is \emph{linear} in $T$.

The above replacement of $\varphi_{0}$ by $\varphi_{C}$ in the supercurrents
is a rather crude approximation as one can see from the inset in the
upper panel of Fig.~\ref{fig:Luitz} where the shape of the zero-temperature
CPR near $\varphi_{C}$ is pretty steep. This limits the validity
of the linear result \eqref{eq:linear_in_T} to very low temperatures
only, which are typically hard to reach by the QMC (see the lower
panel of Fig.~\ref{fig:Luitz}) whose results lie already in the
nonlinear regime (in Ref.~\cite{Delagrange-2015} identified as quadratic).
Extrapolation from that region (e.g., the parabolic fit in Ref.~\cite{Delagrange-2015})
does not respect the true linear low-temperature asymptotics and,
therefore, gives an erroneous estimate as can be seen in the lower
panel of Fig.~\ref{fig:Luitz}. Instead of such highly problematic
and demanding extrapolation procedures (both nonlinear and linear)
we strongly suggest the above crossing of finite-temperature curves
as a simple, robust, and reliable method for determining the position
of the quantum critical point from the finite-temperature data. It
is, moreover, not limited to the phase-dependence of the supercurrent
only, but could be equally used for other measurable quantities as
functions of any experimental control parameter.

\section{Conclusions}

Although currently still in their infancy and predominantly subjects of basic physical research, nanoscopic hybrid devices composed of quantum dots connected to superconducting electrodes are likely to play an important role as functional elements in future electronics technologies. One of the necessary prerequisites for achieving the transfer from fundamental physical understanding to technological applications is the development of efficient and reliable description tools for characterization and simulation of real devices. In particular, in view of today's state-of-the-art of the theoretical description of such systems via heavy (expensive and slow) numerical techniques such as NRG or QMC, the efficiency is a critical issue.          

Our work makes an important step in this direction by offering simple and practical concepts and formulas for the characterization of nanoscopic superconducting hybrids generically described by the superconducting single-impurity Anderson model.    
We have addressed several topics concerning the $0-\text{\ensuremath{\pi}}$
transition both in the ground state and at finite temperatures. 

We have presented two simple analytical formulae \eqref{eq:mGAL} (for
the weakly-correlated regime) and \eqref{eq:Kondo} (for the Kondo
regime) which capture the position of the quantum phase transition
well for a wide range of parameters, especially including away-from-half-filling.
In the cross-over region, where the singlet ground state is neither
purely BCS nor purely Kondo, the equations still provide at least
an estimate for the critical gap (Fig.~1). Despite their approximate
nature these formulas yield correct parametric dependences of the
phase boundary, which is very useful for efficient scans of the parameter
space needed, especially in the initial phase of the data interpretation.

For low-enough finite temperatures, which are nevertheless currently experimentally accessible (below 400~mK), the physics of the
system is governed by the two lowest many-body energy levels, whose
energy difference determines the energy of Andreev bound states. As
a consequence the current-phase relations for different (low-enough)
temperatures cross at a single point, and this crossing marks the
quantum critical point (it should be stressed that this crossing point
is \emph{not} equal to the position where the supercurrent goes through
zero). We propose using this crossing as an easy way how to find
the quantum phase transition directly from finite temperature data.
Moreover, the crossing method is quite universal in that it is not limited to the current-phase relation but
works equally for any other quantity as a function of an arbitrary
control parameter inducing the $0-\pi$ transition (Fig.~\ref{fig:FiniteT_test1}).

Eventually, we have tested the status of the two state-of-the-art numerical methods (NRG and QMC) 
in the context of the experimentally relevant range of parameters of the superconducting single-impurity Anderson model. 
By extensive numerical comparisons we have confirmed the agreement between the NRG and QMC methods and, consequently, their reliability for modeling 
such Josephson junctions in the achievable range of temperatures. Even in their present implementations, they can be safely employed to pinpoint the parameter
values characterizing a given device (optimally after the initial guess is framed by our analytical formulas) and for further simulations of their performance.    

\begin{acknowledgments}
This work was supported by the Czech Science Foundation via Project
No. 16-19640S (T.N., M.\v{Z}., A.K.), the PRIMUS/Sci/09 program of the Charles
University (V.P., A.K.), the Charles University project GA UK No. 888217
(A.K.), National Science Centre (NCN, Poland) via grant number UMO-2017/27/B/ST3/01911 (T.N.), 
and the COST Action NANOCOHYBRI (CA16218) (T.N.). Computational
resources were provided by The Ministry of Education, Youth and Sports
from the Large Infrastructures for Research, Experimental Development
and Innovations project ,,IT4Innovations National Supercomputing Center
-- LM2015070``. Also, access to computing and storage facilities
owned by parties and projects contributing to the National Grid Infrastructure
MetaCentrum provided under the programme \textquotedbl Projects of
Large Research, Development, and Innovations Infrastructures\textquotedbl{}
(CESNET LM2015042) is greatly appreciated. 
\end{acknowledgments}

\appendix*
\section{Modified GAL \label{sec:Appendix}}

In Refs.~\cite{Zonda-2015,Zonda-2016} we obtained an analytical
formula for the $0-\text{\ensuremath{\pi}}$ phase boundary from the
first-order spin-symmetric Hartree-Fock approximation, and noticed
that it's accuracy is significantly improved if the contribution from
the band is neglected. Using the variables $\text{\ensuremath{\chi}}$,
$\tilde{\varepsilon}$ and $\mathcal{U}$ from the main text, the
form was 

\begin{equation}
\chi=\mathcal{U}^{2}-\gamma^{2}\tilde{\varepsilon}^{2}\ ,\label{eq:GAL}
\end{equation}
with the value of the coefficient $\gamma_{\mathrm{GAL}}=U/2\Gamma$.
This formula was called the \emph{generalized atomic limit} (GAL)
\cite[Eq. (17)]{Zonda-2016} in analogy with the atomic limit $(\Delta\rightarrow\infty)$,
where the band is also suppressed, and was found to be a surprisingly
good fit to the NRG data near half filling ($\varepsilon\approx-U/2$,
i.e., $\tilde{\varepsilon}\approx0$), even competing with numerical
results of the second-order diagrammatic approach.

To find a more accurate coefficient $\gamma$ and thus improve the
agreement away from half-filling $\tilde{\varepsilon}\neq0$, we have
plotted the numerical data in an $\tilde{\varepsilon}^{2}-\mathcal{U}$
graph and found that for $\chi=1\,(\varphi=0)$ the dependence is
described by $1-\tilde{\varepsilon}^{2}=1/\mathcal{U}$ for not too
large $\Gamma/\Delta$. Putting this condition into the dependence
\eqref{eq:GAL} with $\gamma$ being now a free parameter, we arrive
at the value $\gamma^{2}=\mathcal{U}(\mathcal{U}+1)$, which
leads to the \textit{Modified} GAL \eqref{eq:mGAL}.


\begin{thebibliography}{56}%
\makeatletter
\providecommand \@ifxundefined [1]{%
 \@ifx{#1\undefined}
}%
\providecommand \@ifnum [1]{%
 \ifnum #1\expandafter \@firstoftwo
 \else \expandafter \@secondoftwo
 \fi
}%
\providecommand \@ifx [1]{%
 \ifx #1\expandafter \@firstoftwo
 \else \expandafter \@secondoftwo
 \fi
}%
\providecommand \natexlab [1]{#1}%
\providecommand \enquote  [1]{``#1''}%
\providecommand \bibnamefont  [1]{#1}%
\providecommand \bibfnamefont [1]{#1}%
\providecommand \citenamefont [1]{#1}%
\providecommand \href@noop [0]{\@secondoftwo}%
\providecommand \href [0]{\begingroup \@sanitize@url \@href}%
\providecommand \@href[1]{\@@startlink{#1}\@@href}%
\providecommand \@@href[1]{\endgroup#1\@@endlink}%
\providecommand \@sanitize@url [0]{\catcode `\\12\catcode `\$12\catcode
  `\&12\catcode `\#12\catcode `\^12\catcode `\_12\catcode `\%12\relax}%
\providecommand \@@startlink[1]{}%
\providecommand \@@endlink[0]{}%
\providecommand \url  [0]{\begingroup\@sanitize@url \@url }%
\providecommand \@url [1]{\endgroup\@href {#1}{\urlprefix }}%
\providecommand \urlprefix  [0]{URL }%
\providecommand \Eprint [0]{\href }%
\providecommand \doibase [0]{http://dx.doi.org/}%
\providecommand \selectlanguage [0]{\@gobble}%
\providecommand \bibinfo  [0]{\@secondoftwo}%
\providecommand \bibfield  [0]{\@secondoftwo}%
\providecommand \translation [1]{[#1]}%
\providecommand \BibitemOpen [0]{}%
\providecommand \bibitemStop [0]{}%
\providecommand \bibitemNoStop [0]{.\EOS\space}%
\providecommand \EOS [0]{\spacefactor3000\relax}%
\providecommand \BibitemShut  [1]{\csname bibitem#1\endcsname}%
\let\auto@bib@innerbib\@empty
\bibitem [{\citenamefont {Mart{\'\i}n-Rodero}\ and\ \citenamefont
  {Levy~Yeyati}(2011)}]{Rodero-2011}%
  \BibitemOpen
  \bibfield  {author} {\bibinfo {author} {\bibfnamefont {A.}~\bibnamefont
  {Mart{\'\i}n-Rodero}}\ and\ \bibinfo {author} {\bibfnamefont
  {A.}~\bibnamefont {Levy~Yeyati}},\ }\href {\doibase
  10.1080/00018732.2011.624266} {\bibfield  {journal} {\bibinfo  {journal}
  {Adv. Phys.}\ }\textbf {\bibinfo {volume} {60}},\ \bibinfo {pages} {899}
  (\bibinfo {year} {2011})}\BibitemShut {NoStop}%
\bibitem [{\citenamefont {De~Franceschi}\ \emph {et~al.}(2010)\citenamefont
  {De~Franceschi}, \citenamefont {Kouwenhoven}, \citenamefont
  {Sch{\"o}nenberger},\ and\ \citenamefont {Wernsdorfer}}]{Wernsdorfer-2010}%
  \BibitemOpen
  \bibfield  {author} {\bibinfo {author} {\bibfnamefont {S.}~\bibnamefont
  {De~Franceschi}}, \bibinfo {author} {\bibfnamefont {L.}~\bibnamefont
  {Kouwenhoven}}, \bibinfo {author} {\bibfnamefont {C.}~\bibnamefont
  {Sch{\"o}nenberger}}, \ and\ \bibinfo {author} {\bibfnamefont
  {W.}~\bibnamefont {Wernsdorfer}},\ }\href
  {http://dx.doi.org/10.1038/nnano.2010.173} {\bibfield  {journal} {\bibinfo
  {journal} {Nat. Nanotechnol.}\ }\textbf {\bibinfo {volume} {5}},\ \bibinfo
  {pages} {703} (\bibinfo {year} {2010})}\BibitemShut {NoStop}%
\bibitem [{\citenamefont {Morpurgo}\ \emph {et~al.}(1999)\citenamefont
  {Morpurgo}, \citenamefont {Kong}, \citenamefont {Marcus},\ and\ \citenamefont
  {Dai}}]{Morpurgo-1999}%
  \BibitemOpen
  \bibfield  {author} {\bibinfo {author} {\bibfnamefont {A.~F.}\ \bibnamefont
  {Morpurgo}}, \bibinfo {author} {\bibfnamefont {J.}~\bibnamefont {Kong}},
  \bibinfo {author} {\bibfnamefont {C.~M.}\ \bibnamefont {Marcus}}, \ and\
  \bibinfo {author} {\bibfnamefont {H.}~\bibnamefont {Dai}},\ }\href
  {http://www.sciencemag.org/content/286/5438/263} {\bibfield  {journal}
  {\bibinfo  {journal} {Science}\ }\textbf {\bibinfo {volume} {286}},\ \bibinfo
  {pages} {263} (\bibinfo {year} {1999})}\BibitemShut {NoStop}%
\bibitem [{\citenamefont {Kasumov}\ \emph {et~al.}(1999)\citenamefont
  {Kasumov}, \citenamefont {Deblock}, \citenamefont {Kociak}, \citenamefont
  {Reulet}, \citenamefont {Bouchiat}, \citenamefont {Khodos}, \citenamefont
  {Gorbatov}, \citenamefont {Volkov}, \citenamefont {Journet},\ and\
  \citenamefont {Burghard}}]{Kasumov-1999}%
  \BibitemOpen
  \bibfield  {author} {\bibinfo {author} {\bibfnamefont {A.~Y.}\ \bibnamefont
  {Kasumov}}, \bibinfo {author} {\bibfnamefont {R.}~\bibnamefont {Deblock}},
  \bibinfo {author} {\bibfnamefont {M.}~\bibnamefont {Kociak}}, \bibinfo
  {author} {\bibfnamefont {B.}~\bibnamefont {Reulet}}, \bibinfo {author}
  {\bibfnamefont {H.}~\bibnamefont {Bouchiat}}, \bibinfo {author}
  {\bibfnamefont {I.~I.}\ \bibnamefont {Khodos}}, \bibinfo {author}
  {\bibfnamefont {Y.~B.}\ \bibnamefont {Gorbatov}}, \bibinfo {author}
  {\bibfnamefont {V.~T.}\ \bibnamefont {Volkov}}, \bibinfo {author}
  {\bibfnamefont {C.}~\bibnamefont {Journet}}, \ and\ \bibinfo {author}
  {\bibfnamefont {M.}~\bibnamefont {Burghard}},\ }\href
  {http://www.sciencemag.org/content/284/5419/1508} {\bibfield  {journal}
  {\bibinfo  {journal} {Science}\ }\textbf {\bibinfo {volume} {284}},\ \bibinfo
  {pages} {1508} (\bibinfo {year} {1999})}\BibitemShut {NoStop}%
\bibitem [{\citenamefont {Kasumov}\ \emph {et~al.}(2003)\citenamefont
  {Kasumov}, \citenamefont {Kociak}, \citenamefont {Ferrier}, \citenamefont
  {Deblock}, \citenamefont {Gu{\'e}ron}, \citenamefont {Reulet}, \citenamefont
  {Khodos}, \citenamefont {St{\'e}phan},\ and\ \citenamefont
  {Bouchiat}}]{Kasumov-2003}%
  \BibitemOpen
  \bibfield  {author} {\bibinfo {author} {\bibfnamefont {A.}~\bibnamefont
  {Kasumov}}, \bibinfo {author} {\bibfnamefont {M.}~\bibnamefont {Kociak}},
  \bibinfo {author} {\bibfnamefont {M.}~\bibnamefont {Ferrier}}, \bibinfo
  {author} {\bibfnamefont {R.}~\bibnamefont {Deblock}}, \bibinfo {author}
  {\bibfnamefont {S.}~\bibnamefont {Gu{\'e}ron}}, \bibinfo {author}
  {\bibfnamefont {B.}~\bibnamefont {Reulet}}, \bibinfo {author} {\bibfnamefont
  {I.}~\bibnamefont {Khodos}}, \bibinfo {author} {\bibfnamefont
  {O.}~\bibnamefont {St{\'e}phan}}, \ and\ \bibinfo {author} {\bibfnamefont
  {H.}~\bibnamefont {Bouchiat}},\ }\href
  {http://link.aps.org/doi/10.1103/PhysRevB.68.214521} {\bibfield  {journal}
  {\bibinfo  {journal} {Phys. Rev. B}\ }\textbf {\bibinfo {volume} {68}},\
  \bibinfo {pages} {214521} (\bibinfo {year} {2003})}\BibitemShut {NoStop}%
\bibitem [{\citenamefont {Jarillo-Herrero}\ \emph {et~al.}(2006)\citenamefont
  {Jarillo-Herrero}, \citenamefont {van Dam},\ and\ \citenamefont
  {Kouwenhoven}}]{Jarillo-2006}%
  \BibitemOpen
  \bibfield  {author} {\bibinfo {author} {\bibfnamefont {P.}~\bibnamefont
  {Jarillo-Herrero}}, \bibinfo {author} {\bibfnamefont {J.~A.}\ \bibnamefont
  {van Dam}}, \ and\ \bibinfo {author} {\bibfnamefont {L.~P.}\ \bibnamefont
  {Kouwenhoven}},\ }\href {http://dx.doi.org/10.1038/nature04550} {\bibfield
  {journal} {\bibinfo  {journal} {Nature}\ }\textbf {\bibinfo {volume} {439}},\
  \bibinfo {pages} {953} (\bibinfo {year} {2006})}\BibitemShut {NoStop}%
\bibitem [{\citenamefont {van Dam}\ \emph {et~al.}(2006)\citenamefont {van
  Dam}, \citenamefont {Nazarov}, \citenamefont {Bakkers}, \citenamefont
  {De~Franceschi},\ and\ \citenamefont {Kouwenhoven}}]{vanDam-2006}%
  \BibitemOpen
  \bibfield  {author} {\bibinfo {author} {\bibfnamefont {J.~A.}\ \bibnamefont
  {van Dam}}, \bibinfo {author} {\bibfnamefont {Y.~V.}\ \bibnamefont
  {Nazarov}}, \bibinfo {author} {\bibfnamefont {E.~P. A.~M.}\ \bibnamefont
  {Bakkers}}, \bibinfo {author} {\bibfnamefont {S.}~\bibnamefont
  {De~Franceschi}}, \ and\ \bibinfo {author} {\bibfnamefont {L.~P.}\
  \bibnamefont {Kouwenhoven}},\ }\href {http://dx.doi.org/10.1038/nature05018}
  {\bibfield  {journal} {\bibinfo  {journal} {Nature}\ }\textbf {\bibinfo
  {volume} {442}},\ \bibinfo {pages} {667} (\bibinfo {year}
  {2006})}\BibitemShut {NoStop}%
\bibitem [{\citenamefont {J{\o}rgensen}\ \emph {et~al.}(2006)\citenamefont
  {J{\o}rgensen}, \citenamefont {Grove-Rasmussen}, \citenamefont {Novotn{\'y}},
  \citenamefont {Flensberg},\ and\ \citenamefont {Lindelof}}]{Jorgensen-2006}%
  \BibitemOpen
  \bibfield  {author} {\bibinfo {author} {\bibfnamefont {H.~I.}\ \bibnamefont
  {J{\o}rgensen}}, \bibinfo {author} {\bibfnamefont {K.}~\bibnamefont
  {Grove-Rasmussen}}, \bibinfo {author} {\bibfnamefont {T.}~\bibnamefont
  {Novotn{\'y}}}, \bibinfo {author} {\bibfnamefont {K.}~\bibnamefont
  {Flensberg}}, \ and\ \bibinfo {author} {\bibfnamefont {P.~E.}\ \bibnamefont
  {Lindelof}},\ }\href {http://link.aps.org/doi/10.1103/PhysRevLett.96.207003}
  {\bibfield  {journal} {\bibinfo  {journal} {Phys. Rev. Lett.}\ }\textbf
  {\bibinfo {volume} {96}},\ \bibinfo {pages} {207003} (\bibinfo {year}
  {2006})}\BibitemShut {NoStop}%
\bibitem [{\citenamefont {Cleuziou}\ \emph {et~al.}(2006)\citenamefont
  {Cleuziou}, \citenamefont {Wernsdorfer}, \citenamefont {Bouchiat},
  \citenamefont {Ondarcuhu},\ and\ \citenamefont {Monthioux}}]{Cleuziou-2006}%
  \BibitemOpen
  \bibfield  {author} {\bibinfo {author} {\bibfnamefont {J.~P.}\ \bibnamefont
  {Cleuziou}}, \bibinfo {author} {\bibfnamefont {W.}~\bibnamefont
  {Wernsdorfer}}, \bibinfo {author} {\bibfnamefont {V.}~\bibnamefont
  {Bouchiat}}, \bibinfo {author} {\bibfnamefont {T.}~\bibnamefont {Ondarcuhu}},
  \ and\ \bibinfo {author} {\bibfnamefont {M.}~\bibnamefont {Monthioux}},\
  }\href {http://dx.doi.org/10.1038/nnano.2006.54} {\bibfield  {journal}
  {\bibinfo  {journal} {Nat. Nanotechnol.}\ }\textbf {\bibinfo {volume} {1}},\
  \bibinfo {pages} {53} (\bibinfo {year} {2006})}\BibitemShut {NoStop}%
\bibitem [{\citenamefont {J{\o}rgensen}\ \emph {et~al.}(2007)\citenamefont
  {J{\o}rgensen}, \citenamefont {Novotn{\'y}}, \citenamefont {Grove-Rasmussen},
  \citenamefont {Flensberg},\ and\ \citenamefont {Lindelof}}]{Jorgensen-2007}%
  \BibitemOpen
  \bibfield  {author} {\bibinfo {author} {\bibfnamefont {H.~I.}\ \bibnamefont
  {J{\o}rgensen}}, \bibinfo {author} {\bibfnamefont {T.}~\bibnamefont
  {Novotn{\'y}}}, \bibinfo {author} {\bibfnamefont {K.}~\bibnamefont
  {Grove-Rasmussen}}, \bibinfo {author} {\bibfnamefont {K.}~\bibnamefont
  {Flensberg}}, \ and\ \bibinfo {author} {\bibfnamefont {P.~E.}\ \bibnamefont
  {Lindelof}},\ }\href {\doibase 10.1021/nl071152w} {\bibfield  {journal}
  {\bibinfo  {journal} {Nano Lett.}\ }\textbf {\bibinfo {volume} {7}},\
  \bibinfo {pages} {2441} (\bibinfo {year} {2007})}\BibitemShut {NoStop}%
\bibitem [{\citenamefont {Grove-Rasmussen}\ \emph {et~al.}(2007)\citenamefont
  {Grove-Rasmussen}, \citenamefont {J{\o}rgensen},\ and\ \citenamefont
  {Lindelof}}]{Grove-2007}%
  \BibitemOpen
  \bibfield  {author} {\bibinfo {author} {\bibfnamefont {K.}~\bibnamefont
  {Grove-Rasmussen}}, \bibinfo {author} {\bibfnamefont {H.~I.}\ \bibnamefont
  {J{\o}rgensen}}, \ and\ \bibinfo {author} {\bibfnamefont {P.~E.}\
  \bibnamefont {Lindelof}},\ }\href
  {http://stacks.iop.org/1367-2630/9/i=5/a=124} {\bibfield  {journal} {\bibinfo
   {journal} {New J. Phys.}\ }\textbf {\bibinfo {volume} {9}},\ \bibinfo
  {pages} {124} (\bibinfo {year} {2007})}\BibitemShut {NoStop}%
\bibitem [{\citenamefont {Pallecchi}\ \emph {et~al.}(2008)\citenamefont
  {Pallecchi}, \citenamefont {Gaass}, \citenamefont {Ryndyk},\ and\
  \citenamefont {Strunk}}]{Pallecchi-2008}%
  \BibitemOpen
  \bibfield  {author} {\bibinfo {author} {\bibfnamefont {E.}~\bibnamefont
  {Pallecchi}}, \bibinfo {author} {\bibfnamefont {M.}~\bibnamefont {Gaass}},
  \bibinfo {author} {\bibfnamefont {D.~A.}\ \bibnamefont {Ryndyk}}, \ and\
  \bibinfo {author} {\bibfnamefont {C.}~\bibnamefont {Strunk}},\ }\href
  {http://link.aip.org/link/?APL/93/072501/1} {\bibfield  {journal} {\bibinfo
  {journal} {Appl. Phys. Lett.}\ }\textbf {\bibinfo {volume} {93}},\ \bibinfo
  {pages} {072501} (\bibinfo {year} {2008})}\BibitemShut {NoStop}%
\bibitem [{\citenamefont {Zhang}\ \emph {et~al.}(8 01)\citenamefont {Zhang},
  \citenamefont {Liu},\ and\ \citenamefont {Lau}}]{Zhang-2008}%
  \BibitemOpen
  \bibfield  {author} {\bibinfo {author} {\bibfnamefont {Y.}~\bibnamefont
  {Zhang}}, \bibinfo {author} {\bibfnamefont {G.}~\bibnamefont {Liu}}, \ and\
  \bibinfo {author} {\bibfnamefont {C.}~\bibnamefont {Lau}},\ }\href {\doibase
  10.1007/s12274-008-8023-6} {\bibfield  {journal} {\bibinfo  {journal} {Nano
  Res.}\ }\textbf {\bibinfo {volume} {1}},\ \bibinfo {pages} {145} (\bibinfo
  {year} {2008-08-01})}\BibitemShut {NoStop}%
\bibitem [{\citenamefont {J{\o}rgensen}\ \emph {et~al.}(2009)\citenamefont
  {J{\o}rgensen}, \citenamefont {Grove-Rasmussen}, \citenamefont {Flensberg},\
  and\ \citenamefont {Lindelof}}]{Jorgensen-2009}%
  \BibitemOpen
  \bibfield  {author} {\bibinfo {author} {\bibfnamefont {H.~I.}\ \bibnamefont
  {J{\o}rgensen}}, \bibinfo {author} {\bibfnamefont {K.}~\bibnamefont
  {Grove-Rasmussen}}, \bibinfo {author} {\bibfnamefont {K.}~\bibnamefont
  {Flensberg}}, \ and\ \bibinfo {author} {\bibfnamefont {P.~E.}\ \bibnamefont
  {Lindelof}},\ }\href {http://link.aps.org/doi/10.1103/PhysRevB.79.155441}
  {\bibfield  {journal} {\bibinfo  {journal} {Phys. Rev. B}\ }\textbf {\bibinfo
  {volume} {79}},\ \bibinfo {pages} {155441} (\bibinfo {year}
  {2009})}\BibitemShut {NoStop}%
\bibitem [{\citenamefont {Liu}\ \emph {et~al.}(2009)\citenamefont {Liu},
  \citenamefont {Zhang},\ and\ \citenamefont {Lau}}]{Liu-2009}%
  \BibitemOpen
  \bibfield  {author} {\bibinfo {author} {\bibfnamefont {G.}~\bibnamefont
  {Liu}}, \bibinfo {author} {\bibfnamefont {Y.}~\bibnamefont {Zhang}}, \ and\
  \bibinfo {author} {\bibfnamefont {C.~N.}\ \bibnamefont {Lau}},\ }\href
  {http://link.aps.org/doi/10.1103/PhysRevLett.102.016803} {\bibfield
  {journal} {\bibinfo  {journal} {Phys. Rev. Lett.}\ }\textbf {\bibinfo
  {volume} {102}},\ \bibinfo {pages} {016803} (\bibinfo {year}
  {2009})}\BibitemShut {NoStop}%
\bibitem [{\citenamefont {Eichler}\ \emph {et~al.}(2009)\citenamefont
  {Eichler}, \citenamefont {Deblock}, \citenamefont {Weiss}, \citenamefont
  {Karrasch}, \citenamefont {Meden}, \citenamefont {Sch{\"o}nenberger},\ and\
  \citenamefont {Bouchiat}}]{Eichler-2009}%
  \BibitemOpen
  \bibfield  {author} {\bibinfo {author} {\bibfnamefont {A.}~\bibnamefont
  {Eichler}}, \bibinfo {author} {\bibfnamefont {R.}~\bibnamefont {Deblock}},
  \bibinfo {author} {\bibfnamefont {M.}~\bibnamefont {Weiss}}, \bibinfo
  {author} {\bibfnamefont {C.}~\bibnamefont {Karrasch}}, \bibinfo {author}
  {\bibfnamefont {V.}~\bibnamefont {Meden}}, \bibinfo {author} {\bibfnamefont
  {C.}~\bibnamefont {Sch{\"o}nenberger}}, \ and\ \bibinfo {author}
  {\bibfnamefont {H.}~\bibnamefont {Bouchiat}},\ }\href
  {http://link.aps.org/doi/10.1103/PhysRevB.79.161407} {\bibfield  {journal}
  {\bibinfo  {journal} {Phys. Rev. B}\ }\textbf {\bibinfo {volume} {79}},\
  \bibinfo {pages} {161407} (\bibinfo {year} {2009})}\BibitemShut {NoStop}%
\bibitem [{\citenamefont {Winkelmann}\ \emph {et~al.}(2009)\citenamefont
  {Winkelmann}, \citenamefont {Roch}, \citenamefont {Wernsdorfer},
  \citenamefont {Bouchiat},\ and\ \citenamefont {Balestro}}]{Winkelmann-2009}%
  \BibitemOpen
  \bibfield  {author} {\bibinfo {author} {\bibfnamefont {C.~B.}\ \bibnamefont
  {Winkelmann}}, \bibinfo {author} {\bibfnamefont {N.}~\bibnamefont {Roch}},
  \bibinfo {author} {\bibfnamefont {W.}~\bibnamefont {Wernsdorfer}}, \bibinfo
  {author} {\bibfnamefont {V.}~\bibnamefont {Bouchiat}}, \ and\ \bibinfo
  {author} {\bibfnamefont {F.}~\bibnamefont {Balestro}},\ }\href {\doibase
  10.1038/nphys1433} {\bibfield  {journal} {\bibinfo  {journal} {Nat. Phys.}\
  }\textbf {\bibinfo {volume} {5}},\ \bibinfo {pages} {876} (\bibinfo {year}
  {2009})}\BibitemShut {NoStop}%
\bibitem [{\citenamefont {Pillet}\ \emph {et~al.}(2010)\citenamefont {Pillet},
  \citenamefont {Quay}, \citenamefont {Morfin}, \citenamefont {Bena},
  \citenamefont {Yeyati},\ and\ \citenamefont {Joyez}}]{Pillet-2010}%
  \BibitemOpen
  \bibfield  {author} {\bibinfo {author} {\bibfnamefont {J.-D.}\ \bibnamefont
  {Pillet}}, \bibinfo {author} {\bibfnamefont {C.~H.~L.}\ \bibnamefont {Quay}},
  \bibinfo {author} {\bibfnamefont {P.}~\bibnamefont {Morfin}}, \bibinfo
  {author} {\bibfnamefont {C.}~\bibnamefont {Bena}}, \bibinfo {author}
  {\bibfnamefont {A.~L.}\ \bibnamefont {Yeyati}}, \ and\ \bibinfo {author}
  {\bibfnamefont {P.}~\bibnamefont {Joyez}},\ }\href
  {http://dx.doi.org/10.1038/nphys1811} {\bibfield  {journal} {\bibinfo
  {journal} {Nat. Phys.}\ }\textbf {\bibinfo {volume} {6}},\ \bibinfo {pages}
  {965} (\bibinfo {year} {2010})}\BibitemShut {NoStop}%
\bibitem [{\citenamefont {Katsaros}\ \emph {et~al.}(2010)\citenamefont
  {Katsaros}, \citenamefont {Spathis}, \citenamefont {Stoffel}, \citenamefont
  {Fournel}, \citenamefont {Mongillo}, \citenamefont {Bouchiat}, \citenamefont
  {Lefloch}, \citenamefont {Rastelli}, \citenamefont {Schmidt},\ and\
  \citenamefont {De~Franceschi}}]{Katsaros-2010}%
  \BibitemOpen
  \bibfield  {author} {\bibinfo {author} {\bibfnamefont {G.}~\bibnamefont
  {Katsaros}}, \bibinfo {author} {\bibfnamefont {P.}~\bibnamefont {Spathis}},
  \bibinfo {author} {\bibfnamefont {M.}~\bibnamefont {Stoffel}}, \bibinfo
  {author} {\bibfnamefont {F.}~\bibnamefont {Fournel}}, \bibinfo {author}
  {\bibfnamefont {M.}~\bibnamefont {Mongillo}}, \bibinfo {author}
  {\bibfnamefont {V.}~\bibnamefont {Bouchiat}}, \bibinfo {author}
  {\bibfnamefont {F.}~\bibnamefont {Lefloch}}, \bibinfo {author} {\bibfnamefont
  {A.}~\bibnamefont {Rastelli}}, \bibinfo {author} {\bibfnamefont
  {O.}~\bibnamefont {Schmidt}}, \ and\ \bibinfo {author} {\bibfnamefont
  {S.}~\bibnamefont {De~Franceschi}},\ }\href {\doibase 10.1038/nnano.2010.84}
  {\bibfield  {journal} {\bibinfo  {journal} {Nat. Nanotechnol.}\ }\textbf
  {\bibinfo {volume} {5}},\ \bibinfo {pages} {458} (\bibinfo {year}
  {2010})}\BibitemShut {NoStop}%
\bibitem [{\citenamefont {Maurand}\ \emph {et~al.}(2012)\citenamefont
  {Maurand}, \citenamefont {Meng}, \citenamefont {Bonet}, \citenamefont
  {Florens}, \citenamefont {Marty},\ and\ \citenamefont
  {Wernsdorfer}}]{Maurand-2012}%
  \BibitemOpen
  \bibfield  {author} {\bibinfo {author} {\bibfnamefont {R.}~\bibnamefont
  {Maurand}}, \bibinfo {author} {\bibfnamefont {T.}~\bibnamefont {Meng}},
  \bibinfo {author} {\bibfnamefont {E.}~\bibnamefont {Bonet}}, \bibinfo
  {author} {\bibfnamefont {S.}~\bibnamefont {Florens}}, \bibinfo {author}
  {\bibfnamefont {L.}~\bibnamefont {Marty}}, \ and\ \bibinfo {author}
  {\bibfnamefont {W.}~\bibnamefont {Wernsdorfer}},\ }\href
  {http://link.aps.org/doi/10.1103/PhysRevX.2.011009} {\bibfield  {journal}
  {\bibinfo  {journal} {Phys. Rev. X}\ }\textbf {\bibinfo {volume} {2}},\
  \bibinfo {pages} {011009} (\bibinfo {year} {2012})}\BibitemShut {NoStop}%
\bibitem [{\citenamefont {Lee}\ \emph {et~al.}(2012)\citenamefont {Lee},
  \citenamefont {Jiang}, \citenamefont {Aguado}, \citenamefont {Katsaros},
  \citenamefont {Lieber},\ and\ \citenamefont {De~Franceschi}}]{Lee-2012}%
  \BibitemOpen
  \bibfield  {author} {\bibinfo {author} {\bibfnamefont {E.~J.~H.}\
  \bibnamefont {Lee}}, \bibinfo {author} {\bibfnamefont {X.}~\bibnamefont
  {Jiang}}, \bibinfo {author} {\bibfnamefont {R.}~\bibnamefont {Aguado}},
  \bibinfo {author} {\bibfnamefont {G.}~\bibnamefont {Katsaros}}, \bibinfo
  {author} {\bibfnamefont {C.~M.}\ \bibnamefont {Lieber}}, \ and\ \bibinfo
  {author} {\bibfnamefont {S.}~\bibnamefont {De~Franceschi}},\ }\href {\doibase
  10.1103/PhysRevLett.109.186802} {\bibfield  {journal} {\bibinfo  {journal}
  {Phys. Rev. Lett.}\ }\textbf {\bibinfo {volume} {109}},\ \bibinfo {pages}
  {186802} (\bibinfo {year} {2012})}\BibitemShut {NoStop}%
\bibitem [{\citenamefont {Pillet}\ \emph {et~al.}(2013)\citenamefont {Pillet},
  \citenamefont {Joyez}, \citenamefont {\v{Z}itko},\ and\ \citenamefont
  {Goffman}}]{Pillet-2013}%
  \BibitemOpen
  \bibfield  {author} {\bibinfo {author} {\bibfnamefont {J.~D.}\ \bibnamefont
  {Pillet}}, \bibinfo {author} {\bibfnamefont {P.}~\bibnamefont {Joyez}},
  \bibinfo {author} {\bibfnamefont {R.}~\bibnamefont {\v{Z}itko}}, \ and\
  \bibinfo {author} {\bibfnamefont {M.~F.}\ \bibnamefont {Goffman}},\ }\href
  {http://link.aps.org/doi/10.1103/PhysRevB.88.045101} {\bibfield  {journal}
  {\bibinfo  {journal} {Phys. Rev. B}\ }\textbf {\bibinfo {volume} {88}},\
  \bibinfo {pages} {045101} (\bibinfo {year} {2013})}\BibitemShut {NoStop}%
\bibitem [{\citenamefont {Kumar}\ \emph {et~al.}(2014)\citenamefont {Kumar},
  \citenamefont {Gaim}, \citenamefont {Steininger}, \citenamefont {Yeyati},
  \citenamefont {Mart\'{i}n-Rodero}, \citenamefont {H\"uttel},\ and\
  \citenamefont {Strunk}}]{Kumar-2014}%
  \BibitemOpen
  \bibfield  {author} {\bibinfo {author} {\bibfnamefont {A.}~\bibnamefont
  {Kumar}}, \bibinfo {author} {\bibfnamefont {M.}~\bibnamefont {Gaim}},
  \bibinfo {author} {\bibfnamefont {D.}~\bibnamefont {Steininger}}, \bibinfo
  {author} {\bibfnamefont {A.~L.}\ \bibnamefont {Yeyati}}, \bibinfo {author}
  {\bibfnamefont {A.}~\bibnamefont {Mart\'{i}n-Rodero}}, \bibinfo {author}
  {\bibfnamefont {A.~K.}\ \bibnamefont {H\"uttel}}, \ and\ \bibinfo {author}
  {\bibfnamefont {C.}~\bibnamefont {Strunk}},\ }\href {\doibase
  10.1103/PhysRevB.89.075428} {\bibfield  {journal} {\bibinfo  {journal} {Phys.
  Rev. B}\ }\textbf {\bibinfo {volume} {89}},\ \bibinfo {pages} {075428}
  (\bibinfo {year} {2014})}\BibitemShut {NoStop}%
\bibitem [{\citenamefont {Delagrange}\ \emph {et~al.}(2015)\citenamefont
  {Delagrange}, \citenamefont {Luitz}, \citenamefont {Weil}, \citenamefont
  {Kasumov}, \citenamefont {Meden}, \citenamefont {Bouchiat},\ and\
  \citenamefont {Deblock}}]{Delagrange-2015}%
  \BibitemOpen
  \bibfield  {author} {\bibinfo {author} {\bibfnamefont {R.}~\bibnamefont
  {Delagrange}}, \bibinfo {author} {\bibfnamefont {D.~J.}\ \bibnamefont
  {Luitz}}, \bibinfo {author} {\bibfnamefont {R.}~\bibnamefont {Weil}},
  \bibinfo {author} {\bibfnamefont {A.}~\bibnamefont {Kasumov}}, \bibinfo
  {author} {\bibfnamefont {V.}~\bibnamefont {Meden}}, \bibinfo {author}
  {\bibfnamefont {H.}~\bibnamefont {Bouchiat}}, \ and\ \bibinfo {author}
  {\bibfnamefont {R.}~\bibnamefont {Deblock}},\ }\href {\doibase
  10.1103/PhysRevB.91.241401} {\bibfield  {journal} {\bibinfo  {journal} {Phys.
  Rev. B}\ }\textbf {\bibinfo {volume} {91}},\ \bibinfo {pages} {241401(R)}
  (\bibinfo {year} {2015})}\BibitemShut {NoStop}%
\bibitem [{\citenamefont {Delagrange}\ \emph {et~al.}(2016)\citenamefont
  {Delagrange}, \citenamefont {Weil}, \citenamefont {Kasumov}, \citenamefont
  {Ferrier}, \citenamefont {Bouchiat},\ and\ \citenamefont
  {Deblock}}]{Delagrange-2016}%
  \BibitemOpen
  \bibfield  {author} {\bibinfo {author} {\bibfnamefont {R.}~\bibnamefont
  {Delagrange}}, \bibinfo {author} {\bibfnamefont {R.}~\bibnamefont {Weil}},
  \bibinfo {author} {\bibfnamefont {A.}~\bibnamefont {Kasumov}}, \bibinfo
  {author} {\bibfnamefont {M.}~\bibnamefont {Ferrier}}, \bibinfo {author}
  {\bibfnamefont {H.}~\bibnamefont {Bouchiat}}, \ and\ \bibinfo {author}
  {\bibfnamefont {R.}~\bibnamefont {Deblock}},\ }\href
  {http://link.aps.org/doi/10.1103/PhysRevB.93.195437} {\bibfield  {journal}
  {\bibinfo  {journal} {Phys. Rev. B}\ }\textbf {\bibinfo {volume} {93}},\
  \bibinfo {pages} {195437} (\bibinfo {year} {2016})}\BibitemShut {NoStop}%
\bibitem [{\citenamefont {Li}\ \emph {et~al.}(2017)\citenamefont {Li},
  \citenamefont {Kang}, \citenamefont {Caroff},\ and\ \citenamefont
  {Xu}}]{Xu-2017}%
  \BibitemOpen
  \bibfield  {author} {\bibinfo {author} {\bibfnamefont {S.}~\bibnamefont
  {Li}}, \bibinfo {author} {\bibfnamefont {N.}~\bibnamefont {Kang}}, \bibinfo
  {author} {\bibfnamefont {P.}~\bibnamefont {Caroff}}, \ and\ \bibinfo {author}
  {\bibfnamefont {H.~Q.}\ \bibnamefont {Xu}},\ }\href {\doibase
  10.1103/PhysRevB.95.014515} {\bibfield  {journal} {\bibinfo  {journal} {Phys.
  Rev. B}\ }\textbf {\bibinfo {volume} {95}},\ \bibinfo {pages} {014515}
  (\bibinfo {year} {2017})}\BibitemShut {NoStop}%
\bibitem [{\citenamefont {Delagrange}\ \emph {et~al.}(2018)\citenamefont
  {Delagrange}, \citenamefont {Weil}, \citenamefont {Kasumov}, \citenamefont
  {Ferrier}, \citenamefont {Bouchiat},\ and\ \citenamefont
  {Deblock}}]{Delagrange-2018-PhysicaB}%
  \BibitemOpen
  \bibfield  {author} {\bibinfo {author} {\bibfnamefont {R.}~\bibnamefont
  {Delagrange}}, \bibinfo {author} {\bibfnamefont {R.}~\bibnamefont {Weil}},
  \bibinfo {author} {\bibfnamefont {A.}~\bibnamefont {Kasumov}}, \bibinfo
  {author} {\bibfnamefont {M.}~\bibnamefont {Ferrier}}, \bibinfo {author}
  {\bibfnamefont {H.}~\bibnamefont {Bouchiat}}, \ and\ \bibinfo {author}
  {\bibfnamefont {R.}~\bibnamefont {Deblock}},\ }\href {\doibase
  https://doi.org/10.1016/j.physb.2017.09.034} {\bibfield  {journal} {\bibinfo
  {journal} {Physica B}\ }\textbf {\bibinfo {volume} {536}},\ \bibinfo {pages}
  {211 } (\bibinfo {year} {2018})}\BibitemShut {NoStop}%
\bibitem [{\citenamefont {Farinacci}\ \emph {et~al.}(2018)\citenamefont
  {Farinacci}, \citenamefont {Ahmadi}, \citenamefont {Reecht}, \citenamefont
  {Ruby}, \citenamefont {Bogdanoff}, \citenamefont {Peters}, \citenamefont
  {Heinrich}, \citenamefont {von Oppen},\ and\ \citenamefont
  {Franke}}]{Farinacci-2018}%
  \BibitemOpen
  \bibfield  {author} {\bibinfo {author} {\bibfnamefont {L.}~\bibnamefont
  {Farinacci}}, \bibinfo {author} {\bibfnamefont {G.}~\bibnamefont {Ahmadi}},
  \bibinfo {author} {\bibfnamefont {G.}~\bibnamefont {Reecht}}, \bibinfo
  {author} {\bibfnamefont {M.}~\bibnamefont {Ruby}}, \bibinfo {author}
  {\bibfnamefont {N.}~\bibnamefont {Bogdanoff}}, \bibinfo {author}
  {\bibfnamefont {O.}~\bibnamefont {Peters}}, \bibinfo {author} {\bibfnamefont
  {B.~W.}\ \bibnamefont {Heinrich}}, \bibinfo {author} {\bibfnamefont
  {F.}~\bibnamefont {von Oppen}}, \ and\ \bibinfo {author} {\bibfnamefont
  {K.~J.}\ \bibnamefont {Franke}},\ }\href {\doibase
  10.1103/PhysRevLett.121.196803} {\bibfield  {journal} {\bibinfo  {journal}
  {Physical Review Letters}\ }\textbf {\bibinfo {volume} {121}},\ \bibinfo
  {pages} {196803} (\bibinfo {year} {2018})}\BibitemShut {NoStop}%
\bibitem [{\citenamefont {Bouchiat}(2009)}]{Bouchiat-2009}%
  \BibitemOpen
  \bibfield  {author} {\bibinfo {author} {\bibfnamefont {V.}~\bibnamefont
  {Bouchiat}},\ }\href {\doibase 10.1088/0953-2048/22/6/064002} {\bibfield
  {journal} {\bibinfo  {journal} {Superconductor Science and Technology}\
  }\textbf {\bibinfo {volume} {22}},\ \bibinfo {pages} {064002} (\bibinfo
  {year} {2009})}\BibitemShut {NoStop}%
\bibitem [{\citenamefont {Luitz}\ \emph {et~al.}(2012)\citenamefont {Luitz},
  \citenamefont {Assaad}, \citenamefont {Novotn{\'y}}, \citenamefont
  {Karrasch},\ and\ \citenamefont {Meden}}]{Luitz-2012}%
  \BibitemOpen
  \bibfield  {author} {\bibinfo {author} {\bibfnamefont {D.~J.}\ \bibnamefont
  {Luitz}}, \bibinfo {author} {\bibfnamefont {F.~F.}\ \bibnamefont {Assaad}},
  \bibinfo {author} {\bibfnamefont {T.}~\bibnamefont {Novotn{\'y}}}, \bibinfo
  {author} {\bibfnamefont {C.}~\bibnamefont {Karrasch}}, \ and\ \bibinfo
  {author} {\bibfnamefont {V.}~\bibnamefont {Meden}},\ }\href
  {http://link.aps.org/doi/10.1103/PhysRevLett.108.227001} {\bibfield
  {journal} {\bibinfo  {journal} {Phys. Rev. Lett.}\ }\textbf {\bibinfo
  {volume} {108}},\ \bibinfo {pages} {227001} (\bibinfo {year}
  {2012})}\BibitemShut {NoStop}%
\bibitem [{\citenamefont {Chang}\ \emph {et~al.}(2013)\citenamefont {Chang},
  \citenamefont {Manucharyan}, \citenamefont {Jespersen}, \citenamefont
  {Nyg{\aa}rd},\ and\ \citenamefont {Marcus}}]{Chang-2013}%
  \BibitemOpen
  \bibfield  {author} {\bibinfo {author} {\bibfnamefont {W.}~\bibnamefont
  {Chang}}, \bibinfo {author} {\bibfnamefont {V.~E.}\ \bibnamefont
  {Manucharyan}}, \bibinfo {author} {\bibfnamefont {T.~S.}\ \bibnamefont
  {Jespersen}}, \bibinfo {author} {\bibfnamefont {J.}~\bibnamefont
  {Nyg{\aa}rd}}, \ and\ \bibinfo {author} {\bibfnamefont {C.~M.}\ \bibnamefont
  {Marcus}},\ }\href {http://link.aps.org/doi/10.1103/PhysRevLett.110.217005}
  {\bibfield  {journal} {\bibinfo  {journal} {Phys. Rev. Lett.}\ }\textbf
  {\bibinfo {volume} {110}},\ \bibinfo {pages} {217005} (\bibinfo {year}
  {2013})}\BibitemShut {NoStop}%
\bibitem [{\citenamefont {Matsuura}(1977)}]{Matsuura-1977}%
  \BibitemOpen
  \bibfield  {author} {\bibinfo {author} {\bibfnamefont {T.}~\bibnamefont
  {Matsuura}},\ }\href {http://ptp.oxfordjournals.org/content/57/6/1823}
  {\bibfield  {journal} {\bibinfo  {journal} {Prog. Theor. Phys.}\ }\textbf
  {\bibinfo {volume} {57}},\ \bibinfo {pages} {1823} (\bibinfo {year}
  {1977})}\BibitemShut {NoStop}%
\bibitem [{\citenamefont {Glazman}\ and\ \citenamefont
  {Matveev}(1989)}]{Glazman-1989}%
  \BibitemOpen
  \bibfield  {author} {\bibinfo {author} {\bibfnamefont {L.~I.}\ \bibnamefont
  {Glazman}}\ and\ \bibinfo {author} {\bibfnamefont {K.~A.}\ \bibnamefont
  {Matveev}},\ }\href@noop {} {\bibfield  {journal} {\bibinfo  {journal} {JETP
  Lett.}\ }\textbf {\bibinfo {volume} {49}},\ \bibinfo {pages} {659} (\bibinfo
  {year} {1989})}\BibitemShut {NoStop}%
\bibitem [{\citenamefont {Rozhkov}\ and\ \citenamefont
  {Arovas}(1999)}]{Rozhkov-1999}%
  \BibitemOpen
  \bibfield  {author} {\bibinfo {author} {\bibfnamefont {A.~V.}\ \bibnamefont
  {Rozhkov}}\ and\ \bibinfo {author} {\bibfnamefont {D.~P.}\ \bibnamefont
  {Arovas}},\ }\href {http://link.aps.org/doi/10.1103/PhysRevLett.82.2788}
  {\bibfield  {journal} {\bibinfo  {journal} {Phys. Rev. Lett.}\ }\textbf
  {\bibinfo {volume} {82}},\ \bibinfo {pages} {2788} (\bibinfo {year}
  {1999})}\BibitemShut {NoStop}%
\bibitem [{\citenamefont {Yoshioka}\ and\ \citenamefont
  {Ohashi}(2000)}]{Yoshioka-2000}%
  \BibitemOpen
  \bibfield  {author} {\bibinfo {author} {\bibfnamefont {T.}~\bibnamefont
  {Yoshioka}}\ and\ \bibinfo {author} {\bibfnamefont {Y.}~\bibnamefont
  {Ohashi}},\ }\href {\doibase 10.1143/JPSJ.69.1812} {\bibfield  {journal}
  {\bibinfo  {journal} {J. Phys. Soc. Jpn.}\ }\textbf {\bibinfo {volume}
  {69}},\ \bibinfo {pages} {1812} (\bibinfo {year} {2000})}\BibitemShut
  {NoStop}%
\bibitem [{\citenamefont {Siano}\ and\ \citenamefont
  {Egger}(2004)}]{Siano-2004}%
  \BibitemOpen
  \bibfield  {author} {\bibinfo {author} {\bibfnamefont {F.}~\bibnamefont
  {Siano}}\ and\ \bibinfo {author} {\bibfnamefont {R.}~\bibnamefont {Egger}},\
  }\href {http://link.aps.org/doi/10.1103/PhysRevLett.93.047002} {\bibfield
  {journal} {\bibinfo  {journal} {Phys. Rev. Lett.}\ }\textbf {\bibinfo
  {volume} {93}},\ \bibinfo {pages} {047002} (\bibinfo {year}
  {2004})}\BibitemShut {NoStop}%
\bibitem [{\citenamefont {Choi}\ \emph {et~al.}(2004)\citenamefont {Choi},
  \citenamefont {Lee}, \citenamefont {Kang},\ and\ \citenamefont
  {Belzig}}]{Choi-2004}%
  \BibitemOpen
  \bibfield  {author} {\bibinfo {author} {\bibfnamefont {M.-S.}\ \bibnamefont
  {Choi}}, \bibinfo {author} {\bibfnamefont {M.}~\bibnamefont {Lee}}, \bibinfo
  {author} {\bibfnamefont {K.}~\bibnamefont {Kang}}, \ and\ \bibinfo {author}
  {\bibfnamefont {W.}~\bibnamefont {Belzig}},\ }\href
  {http://link.aps.org/doi/10.1103/PhysRevB.70.020502} {\bibfield  {journal}
  {\bibinfo  {journal} {Phys. Rev. B}\ }\textbf {\bibinfo {volume} {70}},\
  \bibinfo {pages} {020502} (\bibinfo {year} {2004})}\BibitemShut {NoStop}%
\bibitem [{\citenamefont {Sellier}\ \emph {et~al.}(2005)\citenamefont
  {Sellier}, \citenamefont {Kopp}, \citenamefont {Kroha},\ and\ \citenamefont
  {Barash}}]{Sellier-2005}%
  \BibitemOpen
  \bibfield  {author} {\bibinfo {author} {\bibfnamefont {G.}~\bibnamefont
  {Sellier}}, \bibinfo {author} {\bibfnamefont {T.}~\bibnamefont {Kopp}},
  \bibinfo {author} {\bibfnamefont {J.}~\bibnamefont {Kroha}}, \ and\ \bibinfo
  {author} {\bibfnamefont {Y.~S.}\ \bibnamefont {Barash}},\ }\href
  {http://link.aps.org/doi/10.1103/PhysRevB.72.174502} {\bibfield  {journal}
  {\bibinfo  {journal} {Phys. Rev. B}\ }\textbf {\bibinfo {volume} {72}},\
  \bibinfo {pages} {174502} (\bibinfo {year} {2005})}\BibitemShut {NoStop}%
\bibitem [{\citenamefont {Novotn{\'y}}\ \emph {et~al.}(2005)\citenamefont
  {Novotn{\'y}}, \citenamefont {Rossini},\ and\ \citenamefont
  {Flensberg}}]{Novotny-2005}%
  \BibitemOpen
  \bibfield  {author} {\bibinfo {author} {\bibfnamefont {T.}~\bibnamefont
  {Novotn{\'y}}}, \bibinfo {author} {\bibfnamefont {A.}~\bibnamefont
  {Rossini}}, \ and\ \bibinfo {author} {\bibfnamefont {K.}~\bibnamefont
  {Flensberg}},\ }\href {http://link.aps.org/doi/10.1103/PhysRevB.72.224502}
  {\bibfield  {journal} {\bibinfo  {journal} {Phys. Rev. B}\ }\textbf {\bibinfo
  {volume} {72}},\ \bibinfo {pages} {224502} (\bibinfo {year}
  {2005})}\BibitemShut {NoStop}%
\bibitem [{\citenamefont {Karrasch}\ \emph {et~al.}(2008)\citenamefont
  {Karrasch}, \citenamefont {Oguri},\ and\ \citenamefont
  {Meden}}]{Karrasch-2008}%
  \BibitemOpen
  \bibfield  {author} {\bibinfo {author} {\bibfnamefont {C.}~\bibnamefont
  {Karrasch}}, \bibinfo {author} {\bibfnamefont {A.}~\bibnamefont {Oguri}}, \
  and\ \bibinfo {author} {\bibfnamefont {V.}~\bibnamefont {Meden}},\ }\href
  {http://link.aps.org/doi/10.1103/PhysRevB.77.024517} {\bibfield  {journal}
  {\bibinfo  {journal} {Phys. Rev. B}\ }\textbf {\bibinfo {volume} {77}},\
  \bibinfo {pages} {024517} (\bibinfo {year} {2008})}\BibitemShut {NoStop}%
\bibitem [{\citenamefont {Meng}\ \emph {et~al.}(2009)\citenamefont {Meng},
  \citenamefont {Florens},\ and\ \citenamefont {Simon}}]{Meng-2009}%
  \BibitemOpen
  \bibfield  {author} {\bibinfo {author} {\bibfnamefont {T.}~\bibnamefont
  {Meng}}, \bibinfo {author} {\bibfnamefont {S.}~\bibnamefont {Florens}}, \
  and\ \bibinfo {author} {\bibfnamefont {P.}~\bibnamefont {Simon}},\ }\href
  {http://link.aps.org/doi/10.1103/PhysRevB.79.224521} {\bibfield  {journal}
  {\bibinfo  {journal} {Phys. Rev. B}\ }\textbf {\bibinfo {volume} {79}},\
  \bibinfo {pages} {224521} (\bibinfo {year} {2009})}\BibitemShut {NoStop}%
\bibitem [{\citenamefont {Camjayi}\ \emph {et~al.}(2017)\citenamefont
  {Camjayi}, \citenamefont {Arrachea}, \citenamefont {Aligia},\ and\
  \citenamefont {von Oppen}}]{vonOppen-2017}%
  \BibitemOpen
  \bibfield  {author} {\bibinfo {author} {\bibfnamefont {A.}~\bibnamefont
  {Camjayi}}, \bibinfo {author} {\bibfnamefont {L.}~\bibnamefont {Arrachea}},
  \bibinfo {author} {\bibfnamefont {A.}~\bibnamefont {Aligia}}, \ and\ \bibinfo
  {author} {\bibfnamefont {F.}~\bibnamefont {von Oppen}},\ }\href {\doibase
  10.1103/PhysRevLett.119.046801} {\bibfield  {journal} {\bibinfo  {journal}
  {Physical Review Letters}\ }\textbf {\bibinfo {volume} {119}},\ \bibinfo
  {pages} {046801} (\bibinfo {year} {2017})}\BibitemShut {NoStop}%
\bibitem [{\citenamefont {Kadlecov{\'a}}\ \emph {et~al.}(2017)\citenamefont
  {Kadlecov{\'a}}, \citenamefont {{\v Z}onda},\ and\ \citenamefont
  {Novotn{\'y}}}]{Kadlecova-2017}%
  \BibitemOpen
  \bibfield  {author} {\bibinfo {author} {\bibfnamefont {A.}~\bibnamefont
  {Kadlecov{\'a}}}, \bibinfo {author} {\bibfnamefont {M.}~\bibnamefont {{\v
  Z}onda}}, \ and\ \bibinfo {author} {\bibfnamefont {T.}~\bibnamefont
  {Novotn{\'y}}},\ }\href {\doibase 10.1103/PhysRevB.95.195114} {\bibfield
  {journal} {\bibinfo  {journal} {Physical Review B}\ }\textbf {\bibinfo
  {volume} {95}},\ \bibinfo {pages} {195114} (\bibinfo {year}
  {2017})}\BibitemShut {NoStop}%
\bibitem [{\citenamefont {Wilson}(1975)}]{Wilson-1975}%
  \BibitemOpen
  \bibfield  {author} {\bibinfo {author} {\bibfnamefont {K.~G.}\ \bibnamefont
  {Wilson}},\ }\href {\doibase 10.1103/RevModPhys.47.773} {\bibfield  {journal}
  {\bibinfo  {journal} {Rev. Mod. Phys.}\ }\textbf {\bibinfo {volume} {47}},\
  \bibinfo {pages} {773} (\bibinfo {year} {1975})}\BibitemShut {NoStop}%
\bibitem [{\citenamefont {Haldane}(1978)}]{Haldane-78b}%
  \BibitemOpen
  \bibfield  {author} {\bibinfo {author} {\bibfnamefont {F.~D.~M.}\
  \bibnamefont {Haldane}},\ }\href
  {http://stacks.iop.org/0022-3719/11/i=24/a=030} {\bibfield  {journal}
  {\bibinfo  {journal} {J. Phys. C}\ }\textbf {\bibinfo {volume} {11}},\
  \bibinfo {pages} {5015} (\bibinfo {year} {1978})}\BibitemShut {NoStop}%
\bibitem [{\citenamefont {{\v Z}onda}\ \emph {et~al.}(2015)\citenamefont {{\v
  Z}onda}, \citenamefont {Pokorn\'y}, \citenamefont {Jani\v{s}},\ and\
  \citenamefont {Novotn\'y}}]{Zonda-2015}%
  \BibitemOpen
  \bibfield  {author} {\bibinfo {author} {\bibfnamefont {M.}~\bibnamefont {{\v
  Z}onda}}, \bibinfo {author} {\bibfnamefont {V.}~\bibnamefont {Pokorn\'y}},
  \bibinfo {author} {\bibfnamefont {V.}~\bibnamefont {Jani\v{s}}}, \ and\
  \bibinfo {author} {\bibfnamefont {T.}~\bibnamefont {Novotn\'y}},\ }\href
  {\doibase 10.1038/srep08821} {\bibfield  {journal} {\bibinfo  {journal} {Sci.
  Rep.}\ }\textbf {\bibinfo {volume} {5}},\ \bibinfo {pages} {8821} (\bibinfo
  {year} {2015})}\BibitemShut {NoStop}%
\bibitem [{\citenamefont {{\v Z}onda}\ \emph {et~al.}(2016)\citenamefont {{\v
  Z}onda}, \citenamefont {Pokorn\'y}, \citenamefont {Jani\v{s}},\ and\
  \citenamefont {Novotn\'y}}]{Zonda-2016}%
  \BibitemOpen
  \bibfield  {author} {\bibinfo {author} {\bibfnamefont {M.}~\bibnamefont {{\v
  Z}onda}}, \bibinfo {author} {\bibfnamefont {V.}~\bibnamefont {Pokorn\'y}},
  \bibinfo {author} {\bibfnamefont {V.}~\bibnamefont {Jani\v{s}}}, \ and\
  \bibinfo {author} {\bibfnamefont {T.}~\bibnamefont {Novotn\'y}},\ }\href
  {\doibase 10.1103/PhysRevB.93.024523} {\bibfield  {journal} {\bibinfo
  {journal} {Phys. Rev. B}\ }\textbf {\bibinfo {volume} {93}},\ \bibinfo
  {pages} {024523} (\bibinfo {year} {2016})}\BibitemShut {NoStop}%
\bibitem [{\citenamefont {Pokorn{\'y}}(2016)}]{SQUAD-code}%
  \BibitemOpen
  \bibfield  {author} {\bibinfo {author} {\bibfnamefont {V.}~\bibnamefont
  {Pokorn{\'y}}},\ }\href@noop {} {\enquote {\bibinfo {title} {{SQUAD} -
  second-order perturbation theory solver for a superconducting quantum dot},}\
  } (\bibinfo {year} {2016}),\ \bibinfo {note}
  {github.com/pokornyv/SQUAD}\BibitemShut {NoStop}%
\bibitem [{\citenamefont {{\v Z}itko}(2014)}]{Ljubljana-code}%
  \BibitemOpen
  \bibfield  {author} {\bibinfo {author} {\bibfnamefont {R.}~\bibnamefont {{\v
  Z}itko}},\ }\href@noop {} {\enquote {\bibinfo {title} {{NRG L}jubljana - open
  source numerical renormalization group code},}\ } (\bibinfo {year} {2014}),\
  \bibinfo {note} {nrgljubljana.ijs.si}\BibitemShut {NoStop}%
\bibitem [{\citenamefont {Bauer}\ \emph {et~al.}(2007)\citenamefont {Bauer},
  \citenamefont {Oguri},\ and\ \citenamefont {Hewson}}]{Bauer-2007}%
  \BibitemOpen
  \bibfield  {author} {\bibinfo {author} {\bibfnamefont {J.}~\bibnamefont
  {Bauer}}, \bibinfo {author} {\bibfnamefont {A.}~\bibnamefont {Oguri}}, \ and\
  \bibinfo {author} {\bibfnamefont {A.~C.}\ \bibnamefont {Hewson}},\ }\href
  {http://stacks.iop.org/0953-8984/19/i=48/a=486211} {\bibfield  {journal}
  {\bibinfo  {journal} {J. Phys.: Cond. Mat.}\ }\textbf {\bibinfo {volume}
  {19}},\ \bibinfo {pages} {486211} (\bibinfo {year} {2007})}\BibitemShut
  {NoStop}%
\bibitem [{Note1()}]{Note1}%
  \BibitemOpen
  \bibinfo {note} {Note that the Kondo temperature \protect \textup {\hbox
  {\mathsurround \z@ \protect \normalfont (\ignorespaces \ref {eq:defTk}\unskip
  \@@italiccorr )}} is also an approximation valid around $\protect \tilde
  {\varepsilon }\approx 0$.}\BibitemShut {Stop}%
\bibitem [{\citenamefont {Seth}\ \emph {et~al.}(2016)\citenamefont {Seth},
  \citenamefont {Krivenko}, \citenamefont {Ferrero},\ and\ \citenamefont
  {Parcollet}}]{cthyb2016}%
  \BibitemOpen
  \bibfield  {author} {\bibinfo {author} {\bibfnamefont {P.}~\bibnamefont
  {Seth}}, \bibinfo {author} {\bibfnamefont {I.}~\bibnamefont {Krivenko}},
  \bibinfo {author} {\bibfnamefont {M.}~\bibnamefont {Ferrero}}, \ and\
  \bibinfo {author} {\bibfnamefont {O.}~\bibnamefont {Parcollet}},\ }\href
  {\doibase 10.1016/j.cpc.2015.10.023} {\bibfield  {journal} {\bibinfo
  {journal} {Comput. Phys. Commun.}\ }\textbf {\bibinfo {volume} {200}},\
  \bibinfo {pages} {274} (\bibinfo {year} {2016})}\BibitemShut {NoStop}%
\bibitem [{\citenamefont {Luitz}\ and\ \citenamefont
  {Assaad}(2010)}]{Luitz-2010}%
  \BibitemOpen
  \bibfield  {author} {\bibinfo {author} {\bibfnamefont {D.~J.}\ \bibnamefont
  {Luitz}}\ and\ \bibinfo {author} {\bibfnamefont {F.~F.}\ \bibnamefont
  {Assaad}},\ }\href {http://link.aps.org/doi/10.1103/PhysRevB.81.024509}
  {\bibfield  {journal} {\bibinfo  {journal} {Phys. Rev. B}\ }\textbf {\bibinfo
  {volume} {81}},\ \bibinfo {pages} {024509} (\bibinfo {year}
  {2010})}\BibitemShut {NoStop}%
\bibitem [{\citenamefont {Pokorn{\'y}}\ and\ \citenamefont {{\v
  Z}onda}(2018)}]{Pokorny-2018}%
  \BibitemOpen
  \bibfield  {author} {\bibinfo {author} {\bibfnamefont {V.}~\bibnamefont
  {Pokorn{\'y}}}\ and\ \bibinfo {author} {\bibfnamefont {M.}~\bibnamefont {{\v
  Z}onda}},\ }\href {\doibase https://doi.org/10.1016/j.physb.2017.08.059}
  {\bibfield  {journal} {\bibinfo  {journal} {Physica B}\ }\textbf {\bibinfo
  {volume} {536}},\ \bibinfo {pages} {488 } (\bibinfo {year}
  {2018})}\BibitemShut {NoStop}%
\bibitem [{\citenamefont {Choi}\ \emph {et~al.}(2005)\citenamefont {Choi},
  \citenamefont {Lee}, \citenamefont {Kang},\ and\ \citenamefont
  {Belzig}}]{Choi-2005comm}%
  \BibitemOpen
  \bibfield  {author} {\bibinfo {author} {\bibfnamefont {M.-S.}\ \bibnamefont
  {Choi}}, \bibinfo {author} {\bibfnamefont {M.}~\bibnamefont {Lee}}, \bibinfo
  {author} {\bibfnamefont {K.}~\bibnamefont {Kang}}, \ and\ \bibinfo {author}
  {\bibfnamefont {W.}~\bibnamefont {Belzig}},\ }\href
  {http://link.aps.org/doi/10.1103/PhysRevLett.94.229701} {\bibfield  {journal}
  {\bibinfo  {journal} {Phys. Rev. Lett.}\ }\textbf {\bibinfo {volume} {94}},\
  \bibinfo {pages} {229701} (\bibinfo {year} {2005})}\BibitemShut {NoStop}%
\bibitem [{Note2()}]{Note2}%
  \BibitemOpen
  \bibinfo {note} {The reason why our zero-temperature extrapolation does not
  coincide precisely with the previous calculation (marked by the green arrow)
  is most probably the Fourier fitting which we could avoid.}\BibitemShut
  {Stop}%
\end{thebibliography}
%

\end{document}